\documentclass{article}

\usepackage{amsmath, amsthm, amscd, amsfonts, amssymb, graphicx, color}
\usepackage[bookmarksnumbered, colorlinks, plainpages]{hyperref}
\usepackage{pstricks}
\usepackage{pspicture,pst-all,multido}
\usepackage[utf8]{inputenc}

\theoremstyle{definition}

\theoremstyle{remark}

\numberwithin{equation}{section}

\newcommand{\e}{\operatorname{e}}

\renewcommand{\i}{\operatorname{i}}

\renewcommand{\d}{\operatorname{d}}

\newcommand{\Q}{\mathbb{Q}}
\newcommand{\C}{\mathbb{C}}
\newcommand{\R}{\mathbb{R}}
\newcommand{\N}{\mathbb{N}}

\newcounter{envcount}%

\newenvironment{Rems}%
{\vspace{\bigskipamount}\refstepcounter{envcount}\textbf{(\theenvcount)\enspace Remarks.}}%
  {\vspace{\bigskipamount}}

{\vspace{\bigskipamount}\refstepcounter{envcount}\textbf{(\theenvcount)\enspace Notation.}}%
  {\vspace{\bigskipamount}}

{\vspace{\bigskipamount}\refstepcounter{envcount}\textbf{(\theenvcount)\enspace Limit hyperplanes.}}%
  {\vspace{\bigskipamount}}

{\vspace{\bigskipamount}\refstepcounter{envcount}\textbf{(\theenvcount)\enspace Remark on}}%
  {\vspace{\bigskipamount}}

\newenvironment{Rem}%
{\vspace{\bigskipamount}\refstepcounter{envcount}\textbf{(\theenvcount)\enspace Remark.}}%
  {\vspace{\bigskipamount}}

\newenvironment{Propa}%
{\vspace{\bigskipamount}\refstepcounter{envcount}\textbf{(\theenvcount)\enspace Properties of achronal sets.}}%
  {\vspace{\bigskipamount}}

  \newenvironment{Propb}%
{\vspace{\bigskipamount}\refstepcounter{envcount}\textbf{(\theenvcount)\enspace Properties of causal bases.}}%
  {\vspace{\bigskipamount}}

\newenvironment{Def}%
{\vspace{\bigskipamount}\refstepcounter{envcount}\textbf{(\theenvcount)\enspace Definition.}}%
  {\vspace{\bigskipamount}}

{\vspace{\bigskipamount}\refstepcounter{envcount}\textbf{(\theenvcount)\enspace POL for Weyl particles.}}%
  {\vspace{\bigskipamount}}

{\vspace{\bigskipamount}\refstepcounter{envcount}\textbf{(\theenvcount)\enspace Summary.}}%
  {\vspace{\bigskipamount}}

{\vspace{\bigskipamount}\refstepcounter{envcount}\textbf{(\theenvcount)\enspace Time Reversal.}}%
  {\vspace{\bigskipamount}}
  
\newenvironment{L-CS}%
{\vspace{\bigskipamount}\refstepcounter{envcount}\textbf{(\theenvcount)\enspace Large-\,$t_{\overline{e}}$\,-states.}}%
  {\vspace{\bigskipamount}}

{\vspace{\bigskipamount}\refstepcounter{envcount}\textbf{(\theenvcount)\enspace Asymptotic causality.}}%
  {\vspace{\bigskipamount}}

\newenvironment{Exa}%
{\vspace{\bigskipamount}\refstepcounter{envcount}\textbf{(\theenvcount)\enspace Example.}}%
  {\vspace{\bigskipamount}}

{\vspace{\bigskipamount}\refstepcounter{envcount}\textbf{(\theenvcount)\enspace Example}}%
  {\vspace{\bigskipamount}}

{\vspace{\bigskipamount}\refstepcounter{envcount}\textbf{(\theenvcount)\enspace Construction}}%
  {\vspace{\bigskipamount}}

{\vspace{\bigskipamount}\refstepcounter{envcount}\textbf{(\theenvcount)\enspace Discussion. }}%
  {\vspace{\bigskipamount}}

\newenvironment{The}%
{\vspace{\bigskipamount}\refstepcounter{envcount}\textbf{(\theenvcount)\enspace Theorem.}\itshape}%
  {\vspace{\bigskipamount}\upshape}
  
\newenvironment{Theo}%
{\vspace{\bigskipamount}\refstepcounter{envcount}\textbf{(\theenvcount)\enspace Theorem}\itshape}%
  {\vspace{\bigskipamount}\upshape}
  
\newenvironment{Pro}%
{\vspace{\bigskipamount}\refstepcounter{envcount}\textbf{(\theenvcount)\enspace Proposition.}\itshape}%
  {\vspace{\bigskipamount}\upshape}

\newenvironment{Cor}%
{\vspace{\bigskipamount}\refstepcounter{envcount}\textbf{(\theenvcount)\enspace Corollary.}\itshape}%
  {\vspace{\bigskipamount}\upshape}

{\vspace{\bigskipamount}\refstepcounter{envcount}\textbf{(\theenvcount)\enspace Frame-dependence of NWL.}\itshape}%
  {\vspace{\bigskipamount}\upshape}

{\vspace{\bigskipamount}\refstepcounter{envcount}\textbf{(\theenvcount)\enspace Construction of a late-change state.}\itshape}%
  {\vspace{\bigskipamount}\upshape}

\newenvironment{Lem}%
{\vspace{\bigskipamount}\refstepcounter{envcount}\textbf{(\theenvcount)\enspace Lemma.}\itshape}%
  {\vspace{\bigskipamount}\upshape}

 \newenvironment{Lemm}%
{\vspace{\bigskipamount}\refstepcounter{envcount}\textbf{(\theenvcount)\enspace Lemma}\itshape}%
  {\vspace{\bigskipamount}\upshape}

{\vspace{\bigskipamount}\refstepcounter{envcount}\textbf{(\theenvcount)\enspace Example.}\itshape}%
  {\vspace{\bigskipamount}\upshape}

{\vspace{\bigskipamount}\refstepcounter{envcount}\textbf{(\theenvcount)\enspace Example}\itshape}%
  {\vspace{\bigskipamount}\upshape}

  {\vspace{\bigskipamount}}

{\vspace{\bigskipamount}\refstepcounter{envcount}\textbf{(\theenvcount)\enspace Interpretation.}\itshape}%
  {\vspace{\bigskipamount}\upshape}

{\vspace{\bigskipamount}\refstepcounter{envcount}\textbf{(\theenvcount)\enspace Canonical cross section.}}%
  {\vspace{\bigskipamount}}

\theoremstyle{definition}
\swapnumbers

\theoremstyle{remark}

\begin{document}
\setcounter{page}{1}
\pagenumbering{arabic}

\vspace{2mm}
\begin{center}
{\Large Achronal Localization, Representations of the Causal Logic for massive systems}\\  

\vspace{1cm}
Domenico P.L. Castrigiano\\
Technische Universit\"at M\"unchen, Fakult\"at f\"ur Mathematik, M\"unchen, Germany\\ 

\smallskip

{\it E-mail address}: {\tt
castrig\,\textrm{@}\,ma.tum.de}\\
ORCID 0000-0002-9819-3212
\end{center}

\abstract 
On plain physical grounds localization of relativistic quantum particles is extended to the achronal regions of Minkowski spacetime. Achronal localization fulfills automatically the requirements of causality. It constitutes the frame which complies most completely with  the principle of causality for quantum mechanical systems. Achronal localization
 is equivalent to the localization in the regions of the causal logic. Covariant representations of the causal logic are constructed for the systems with mass spectrum of positive Lebesgue measure and  every definite spin. Apparently  no representation of the causal logic  for a real mass system has been achieved in the past.

\section{Introduction}

Heuristically, in relativistic quantum physics for free particles, causality has a simple meaning. Particles do not move faster than light. It is not easy to take account of this fundamental fact.  It excludes the description of position following Newton, Wigner and Wightman \cite{NWWH} viz. by a projection operator valued normalized measure. As worked out by Hegerfeldt \cite{NWWH},
positivity of energy  brings about  that a particle initially  localized in a bounded region would jump instantaneously everywhere.
In addition a Wightman localization does not even exist for massless particles with non-zero helicity.  
\\
\hspace*{6mm}
\textbf{Causality.} 
So one  renounces  the possibility of  localized states  but considers instead a regional probability of localization. Causality requires that the probability of localization in a region of influence  determined by the limiting velocity of light  is not less than that in the region of actual localization. The probability of localization in a bounded region is  less than $1$ and no measurement of position can change that. 
\\
\hspace*{6mm}
We do not enter the discussion whether this reveals a violation of the principle of locality but take here a pragmatic stance. For an early contribution see \cite{G70}.
\\
\hspace*{6mm}
In principle one assumes  the probability of localization in a region 
 to be the expectation value of a positive localization operator  assigned to that region.  Position is described by a positive operator valued normalized measure (POM).  POM for massless particles  do exist \cite{C81}. POM on flat spacelike regions which  comply with causality (\ref{CCFR}) are known for the Dirac and Weyl fermions \cite{C17} and for the massive scalar boson (see below).
 \\
\hspace*{6mm}
\textbf{Achronal localization.}
Concerning the regions of localization there are plain physical grounds why one must not restrict oneself to regions of Euclidean space.
One had to deal with a rather general class of \textbf{spacetime regions} in Minkowski space. This class has to be (i) relativistic invariant and (ii) closed under unions of countably many spacelike separated elements. 
Furthermore, as argued in sec.\,\ref{UEE}, it (iii) contains the  \textbf{high boost limits} of spacelike hyperplanes, i.e., the tangent spaces of the light cones.
 \\
\hspace*{6mm}
The classes we identify are either constituted by the achronal Borel sets of spacetime $\mathcal{B}^{ach}$   or by the sets of the causal logic $\mathcal{C}$.   \textbf{Achronal localization} (AL) and \textbf{representation of the causal logic} (RCL) denote the respective localizations by POM. AL  automatically fulfills causality (\ref{CCPOB}).
Actually  it turns out, and this is an important result (\ref{AC}), that these approaches are equivalent. We speak of achronal localization as such. It is the frame which complies most completely with  the principle of causality for quantum mechanical systems.
 \\
\hspace*{6mm}
\textbf{Causal density currents.}  
Only recently considerable progress in extending localization to spacetime regions has been achieved. It mainly concerns causal  localizations for the massive scalar boson and refers to the important work of Petzold and collaborators \cite{GGP67}, \cite{GGPR68}, where apparently  for the first time the conserved covariant four-vector  density currents with a positive definite zeroth component are studied.  \cite{GGP67}  and subsequently  \cite{HW71} reveal the basic one-parameter family of causal kernels of the density currents. The causal kernels are thoroughly analyzed in \cite{C23}. Let us emphasize  that they determine the only concrete causal density currents known up to now.
 \\
\hspace*{6mm}
For the corresponding localizations obtained by integrating the zeroth component   over regions of Euclidean space it is  shown in \cite{GGPR68}  that the temporal evolution is causal. Within the frame of flat spacetime regions, in \cite{C23} a rigorous proof of (full) causality is given adapting  a method in \cite{M23}. De Rosa and Moretti \cite{DM24} considerably extend this result for POM on smooth Cauchy surfaces improving further the method of integration of the causal density current.  
 \\
\hspace*{6mm}
 \textbf{Representation of the causal logic.}  
The class of Borel subsets of smooth Cauchy surfaces is vast but still not closed under unions of spacelike separated elements so that it does not meet completely the requirements of causality. Take it for the moment that one succeeds in integrating the causal density current over all achronal set. A promising  method is applied in \cite{CJ79}. The result is an AL.  In case of the massive scalar boson it is even Poincar\'e covariant and gives rise to the equivalent Poincar\'e covariant RCL.
 \\
\hspace*{6mm} 
Despite of its physical relevance apparently  no representation of the causal logic  for a real mass system has been achieved so far. In the literature \cite{BJ79}, 
\cite{CJ79},
\cite{CJn77} the approach is via a causal density current  (also  operator valued) as illustrated above. There are the problems to find the causal density currents, which are covariant according to  the physical system under consideration and to integrate them over every achronal Borel set.
 \\
\hspace*{6mm}
\textbf{Group representation theoretical approach.} We succeed pursuing a different approach.
The basic ingredient is the natural action of the Poincar\'e group on the set of timelike straight lines. This idea is due to Doplicher, Regge, Singer 1968 \cite{DRS68}, who invented a model showing the independence of locality and positivity of the energy.  Then the canonical system of imprimitivities furnishes a huge representation of the 
Poincar\'e group containing subrepresentations for every mass spectrum of positive Lebesgue measure and for every spin. Assigning to every set of the causal logic the canonical projection, respectively its trace on the invariant subspace, for the set of their emitted lines determines a covariant RCL. See the theorem (\ref{CANCLL}) and the main result (\ref{MT}).

\section{Preliminaries}
Here  general notations and rather detailed the basic notions follow.

\subsection{Notations} 
Vectors in $\R^4$ are denoted  by  $\mathfrak{x}=(x_0,x)$ with $x:=(x_1,x_2,x_3)\in\R^3$. Let $\varpi:\R^4\to\R^3$ denote the projection $\varpi(\mathfrak{x}):=x$.  Representing Minkowski spacetime by $\R^4$ the Minkowski product of $\mathfrak{a},\mathfrak{a}'\in\R^4$ is given by $\mathfrak{a}\cdot \mathfrak{a}':=a_0a_0'-aa':=a_0a'_0-a_1a'_1-a_2a'_2-a_3a'_3$. Often we use the notation $\mathfrak{a}^{\cdot 2}:=\mathfrak{a}\cdot\mathfrak{a}$. 
\\
\hspace*{6mm}
$\tilde{\mathcal{E}}=ISU(2)$  and  $\tilde{\mathcal{P}}=ISL(2,\C)$  are the universal covering groups 
of the Euclidean group and the Poincar\'e group, respectively.     
$\tilde{\mathcal{P}}$ acts on $\R^4$ as
\begin{equation}\label{PTUCH} 
g\cdot \mathfrak{x}:=\mathfrak{a}+\varLambda(A) \mathfrak{x}\quad  \text{ for } g=(\mathfrak{a},A)\in\tilde{\mathcal{P}}, \, \mathfrak{x}\in \R^4
\end{equation}
where $\varLambda:SL(2,\C)\to O(1,3)_0$ is  the universal covering homomorphism onto the proper orthochronous Lorentz group. For short one writes  $A\equiv (0,A)$, 
$\mathfrak{a} \equiv (\mathfrak{a},I_2)$, and $A\cdot \mathfrak{x}=\varLambda(A)\mathfrak{x}$. 
For $M\subset\R^4$ and $g\in \tilde{\mathcal{P}}$ define $g\cdot M:=\{g\cdot \mathfrak{x}: \mathfrak{x}\in M\}$.\\
\hspace*{6mm}
The group operation on  $\tilde{\mathcal{P}}$ reads 
$(\mathfrak{a},A)(\mathfrak{a}',A')
=(\mathfrak{a}+A\cdot\mathfrak{a}',AA')$ with identity element $(0,I_2)$ and inverse  $(\mathfrak{a},A)^{-1}=(-A^{-1}\cdot \mathfrak{a},A^{-1})$.\\
 \hspace*{6mm}
 Identifying $\Lambda(SU(2))\equiv SO(3)$, $SU(2)$ acts on $\R^3$. The elements of $\tilde{\mathcal{E}}$ often are denoted by $(b,B)$ with $b\in\R^3$.

\subsection{Achronal sets, causal bases, causal logic}

 A vector $\mathfrak{z}\in\R^4$ is called spacelike, timelike, or lightlike  if $\mathfrak{z}^{\cdot 2}$ is $<0$,  $>0$,  or $=0$, respectively. Furthermore $\mathfrak{z}\in\R^4$ is \textbf{achronal} if $\mathfrak{z}^{\cdot 2}\le 0$ and   \textbf{causal} if $\mathfrak{z}^{\cdot 2}\ge 0$.\\ 
 \hspace*{6mm}
  A  line\footnote{line is short for straight line} $\mathfrak{a}+\R\mathfrak{z}$, where $\mathfrak{a}\in\R^4$, $\mathfrak{z}\ne 0$,  is spacelike, timelike, etc., if so is $\mathfrak{z}$.
 \\
  \hspace*{6mm}
 A set $\Delta\subset\R^4$ is \textbf{achronal} if every $\mathfrak{x}$, $\mathfrak{y}\in\Delta$ are  \textbf{achronal separated}, i.e., $(\mathfrak{x}-\mathfrak{y})^{\cdot 2} \le 0$. It is 
 \textbf{spacelike}   if any two different points $\mathfrak{x}$, $\mathfrak{y}\in\Delta$ are \textbf{spacelike separated}, i.e., $(\mathfrak{x}-\mathfrak{y})^{\cdot 2} <0$. \\
 \hspace*{6mm}  
  Two  sets $\Delta,\Gamma\subset \R^4$ are said to be \textbf{achronal separated} if they are disjoint and  $(\mathfrak{x}-\mathfrak{y})^{\cdot 2}\le 0$ for $\mathfrak{x}\in\Delta$ and $\mathfrak{y}\in\Gamma$. Similarly they are called  \textbf{spacelike separated} or \textbf{causally independent} if   $(\mathfrak{x}-\mathfrak{y})^{\cdot 2}<0$ for $\mathfrak{x}\in\Delta$ and $\mathfrak{y}\in\Gamma$. \\
\hspace*{6mm}
 Two events $\mathfrak{x},\mathfrak{y}$ are said  to be 
\textbf{ timelike separated} if $(\mathfrak{x}-\mathfrak{y})^{\cdot 2}>0$ and \textbf{ lightlike separated} if $(\mathfrak{x}-\mathfrak{y})^{\cdot 2}=0$.
\\
\hspace*{6mm}
By definition, a \textbf{maximal achronal set} is
 an achronal set, which  it is not properly contained  in an achronal set.  \textbf{Maximal spacelike sets} are analogous.

\subsubsection{Achronal sets}
 In the following several properties of achronal sets are collected. Frequently we shall use (a), (c), (d), (e).  The decisive result (e) is due to \cite[Corollary 1]{CJ77}.

\begin{Propa} \label{PAS} \\

(a) {\it A set is achronal if  and only if it meets every timelike line  at most once. }
\\
\hspace*{6mm}

{\it Proof.}  Let $A$ be achronal and assume that the timelike line $\mathfrak{a}+\R\mathfrak{z}$ meets $A$ twice. Then   $\mathfrak{a}+s\mathfrak{z}$,  $\mathfrak{a}+s'\mathfrak{z}$ lie in $A$ for some $s\ne s'$. This is impossible as  $\mathfrak{a}+s\mathfrak{z}$,  $\mathfrak{a}+s'\mathfrak{z}$ are timelike separated. Conversely, assume  that $A$ is not achronal. Then there are timelike separated $\mathfrak{x}$,  $\mathfrak{y}$ in $A$, and the timelike line $\mathfrak{x}+\R(\mathfrak{y}-\mathfrak{x})$ meets $A$ twice, which is excluded.\qed
\\

(b) {\it Let $A$ be an achronal set. Then there is a unique function $\tau:\varpi(A)\to \R$ such that $A=\{(\tau(x),x):x\in\varpi(A)\}$.   It is $1$-Lipschitz on $\varpi(A)$, i.e., $|\tau(x)-\tau(x)|\le |x-y|$, $x,y\in\varpi(A)$.   Conversely, if $\Delta\subset\R^3$ and $\tau$ is  $1$-Lipschitz on $\Delta$, then   $A:=\{(\tau(x),x):x\in\Delta\}$ is achronal.} 
\\
\hspace*{6mm}

{\it Proof.} Let $A$ be achronal.  By (a), for $x\in\varpi(A)$ the timelike line $(0,x)+\R(1,0,0,0)$  meets $A$, actually only once. In other words  for $\mathfrak{x}\in A$ there is a unique $\tau(x)\in\R$ with $(0,x)+\tau(x)(1,0,0,0)=(\tau(x),x)\in A$, whence $x_0=\tau(x)$. $\tau:\varpi(A)\to \R$ is $1$-Lipschitz as $A$ is achronal. The converse is obvious.\qed
\\

(c) {\it  Every achronal set is contained in a maximal one. More generally, let $M\subset\R^4$. Then every achronal set $A\subset M$ is contained in a maximal achronal set in $M$}. 
\\
\hspace*{6mm}

{\it Proof.}  This holds  true by an obvious application of Zorn's Lemma. Indeed, let $A\subset M$ be an achronal set and denote by $ \mathcal{A}$ the set of all achronal sets in $M$ containing $A$. It is partially ordered by the ordinary set inclusion. Let $\mathcal{C}\subset\mathcal{A}$ be a chain, put $\hat{C}:=\bigcup \mathcal{C}\subset M$, and consider $\mathfrak{x},\mathfrak{x}'\in \hat{C}$. There are $C,C'\in\mathcal{C}$ with $\mathfrak{x}\in C$,  $\mathfrak{x}'\in C'$. Without restriction $C\subset C'$ and  $\mathfrak{x}\ne\mathfrak{x}$. Hence 
 $\mathfrak{x},\mathfrak{x}'\in C'$ are achronal separated. Therefore $\hat{C}\in\mathcal{A}$ and is an upper bound of $\mathcal{C}$. Thus $\mathcal{A}$ has a maximal element $\hat{A}$. $\hat{A}$ is maximal achronal and contains $A$.\qed
\\

(d) {\it The closure of an achronal set is achronal. Hence every maximal achronal set $A$ is a closed set}. 
\\
\hspace*{6mm}

{\it Proof.}  Let $\mathfrak{x}_n, \mathfrak{x}'_n\in A$ with 
$\mathfrak{x}_n\to \mathfrak{x}$ and 
$\mathfrak{x}'_n\to \mathfrak{x}'$ for some $\mathfrak{x},\mathfrak{x}'\in\R^4$. Then $0\ge (\mathfrak{x}_n-\mathfrak{y})^{\cdot 2}\to (\mathfrak{x}-\mathfrak{y})^{\cdot 2}$
for all $\mathfrak{y}\in A$, and $0\ge (\mathfrak{x}_n-\mathfrak{x}'_n)^{\cdot 2}\to (\mathfrak{x}-\mathfrak{x}')^{\cdot 2}$, whence the claim.\qed
\\

(e)  {\it An achronal set  is maximal achronal if and only if it meets every timelike  line.}
\\
\hspace*{6mm}

{\it Proof.} By \cite[Corollary 1]{CJ77} a maximal achronal set meets every timelike line. Conversely, let $A$ be achronal meeting all timelike lines. Assume that $A$ is  not maximal achronal. Then there is $\mathfrak{a}\in\R^4\setminus A$ with $(\mathfrak{x}-\mathfrak{a})^{\cdot 2}\le 0$ for all $\mathfrak{x}\in A$. Since $\mathfrak{a}+\R(1,0,0,0)$ is timelike, there is $s\ne 0$ with $\mathfrak{y}:=\mathfrak{a}+s(1,0,0,0) \in A$. However $(\mathfrak{y}-\mathfrak{a})^{\cdot 2}>0$, which is a contradiction.\qed
\\ 
 
(f) {\it An  achronal set  $A$ is maximal achronal if and only if $\varpi(A)=\R^3$.}
\\
\hspace*{6mm}

{\it Proof.} Let $A$ be maximal achronal. Then $A\cap \big((0,x)+\R(1,0,0,0)\big) \ne \emptyset$  for every $x\in\R^3$ by (e), whence the claim. Conversely, let $\mathfrak{y}\not\in A$. Then by assumption there is $\mathfrak{x}\in A$ with $x=y$ and hence $(\mathfrak{x}-\mathfrak{y})^{\cdot 2}=(x_0-y_0)^2>0$ finishing the proof.\qed
\\

(g) {\it A set  $A$ is maximal achronal if and only if  $A=\{\big(\tau(x),x\big):x\in\R^3\}$    for a $1$-Lipschitz  function $\tau:\R^3\to \R$. $\tau$ is uniquely determined.  }
\\
\hspace*{6mm}

{\it Proof.} 
 See \cite[(2)]{CJ79}. It is an  corollary of \cite[Corollary 1]{CJ77}, see (e).\qed
\\ 
 
(h) {\it Let $A=\{\big(\tau(x),x\big):x\in\R^3\}$ from \emph{(g)} be endowed with the induced topology from $\R^4$.   Then $j_A:\R^3\to A$, $j_A(x):=(\tau(x),x)$  is a homeomorphism, $\{\mathfrak{x}\in\R^4: x_0<\tau(x)\}$ and  $\{\mathfrak{x}\in\R^4: x_0>\tau(x)\}$ are open with  boundary $A$, and  $A$ is path-connected.}
\\
\hspace*{6mm}

{\it Proof.}     The claims are  rather obvious.\qed
\\

(i)  {\it A set  is spacelike maximal achronal if and only if  $|\tau(x)-\tau(y)|< |x-y|$, $x \ne y$  for $\tau$ from} (g). 
\\
\hspace*{6mm}

{\it Proof.} This follows immediately from (g).\qed

\end{Propa}

\begin{Rem}\label{AGSTT}
The term achronal (separated) is used in general spacetime theories \cite{O83} meaning that two events $\mathfrak{x}\ne \mathfrak{y}$ cannot be joined by a timelike $C^1$-curve  $\gamma$. Actually this is an equivalent definition. Indeed, it  implies that the line joining  $\mathfrak{x}$ and  $\mathfrak{y}$ is not timelike, whence $(\mathfrak{x}-\mathfrak{y})^{\cdot 2} \le 0$.
The converse holds, too. Write $\gamma(s)=:(c(s),g(s))$. By timelikeness $|\dot{c}(s)|>|\dot{g}(s)|$. Hence for $s<s'$ one has $|g(s')-g(s)|=|\int_s^{s'} \dot{g}(\alpha)\d \alpha|\le \int_s^{s'}|\dot{g}(\alpha)|\d \alpha < \int_s^{s'}|\dot{c}(\alpha)|\d \alpha=|c(s')-c(s)|$ by the uniform sign of $\dot{c}(s)$ due to Bolzano's theorem. Therefore $\gamma(s')-\gamma(s)$ is timelike, whence the claim.\\
\hspace*{6mm}
For later use we  complete the above result.  Recall that a curve $\gamma:]a,b[\to\R^4$ for $\infty\le a<b\le\infty$ is called \textbf{inextendible} if neither $\lim_{s\to a} \gamma(s)$ nor $\lim_{s\to b} \gamma(s)$ exists.
\\
\hspace*{6mm}
 {\it Let the $C^1$-curve $\gamma$ be timelike, viz.,  $\dot{\gamma}(s)$  timelike for all $s$. 
Then $\gamma(s')-\gamma(s)$ is timelike for all $s\ne s'$ and $\gamma_0$ is strictly monotone. If $\gamma$ is inextendible, then $|\gamma_0(s)|\to \infty$ for $s\to a$ and $s\to b$.}

{\it Proof.} Put $\gamma=(c,g)$. As shown, $|g(s')-g(s)|<|c(s')-c(s)|$ for $s\ne s'$ and, since the sign of $\dot{c}$  is uniform, $c$ is strictly monotone. Now let $\gamma$ be inextendible. Assume that $c$ is bounded above. Without restriction $c(s)\to \infty$ for $s\to b$. By monotony $c(s)\to \sup c$ for $s\to b$. Hence $|g(s')-g(s)|\le |c(s')-c(s)|\to 0$ for  $s,s'\to b$. Then by Cauchy's criterion for sequences $\lim_{s\to b} g(s)$ exists, contradicting inextendibility of $\gamma$. Therefore $c$ is not bounded above. Similarly $c$ is not bounded below.\qed
\end{Rem}\\
Obviously, spacelike sets  are achronal. However there are achronal sets which are not  spacelike. Important examples are the  achronal  not spacelike hyperplanes. These are just  the tangent spaces of the light cones. 

\begin{Exa}\label{SNSAH}
{\it The set $\chi:=\{\mathfrak{x}\in\R^4:x_0=x_3\}$ is } (a) {\it a hyperplane,} (b) {\it  achronal, } (c)  {\it  not spacelike,} (d)   {\it   maximal achronal.} (e) {\it Every timelike line intersects $\chi$.} (f) {\it  $\gamma(t):=\big(t,0,0,\frac{1}{2}(\sqrt{1+t^2}+t)\big)$, $t\in\R$, is an inextendible  timelike smooth curve, which does not  intersect $\chi$.} (g) {\it A lightlike line $\mathfrak{a}+\R\mathfrak{z}$ meets $\chi$ if and only if $\mathfrak{a}\in\chi$ or  $\mathfrak{z}\not\in\chi$.} 
\end{Exa}\\
{\it Proof.} (a) holds, since $\chi=\R(1,0,0,1)+\R(0,1,0,0)+\R(0,0,1,0)$. (b) For $\mathfrak{x},\mathfrak{y}\in\chi$ one has  $(\mathfrak{x}-\mathfrak{y})^{\cdot 2}=-(x_1-y_1)^2-(x_2-y_2)^2\le0$. (c) $\mathfrak{x}:=(1,0,0,1)$ and   $\mathfrak{y}:=0$ are points of $\chi$, which are lightlike related. (d) Suppose that $\mathfrak{y}\in\R^4$ is achronal to all $\mathfrak{x}\in\chi$. Choose $\mathfrak{x}=(y_0,y_1,y_2,y_0)\in\chi$. Then $(\mathfrak{y}-\mathfrak{x})^{\cdot 2}=(y_0-x_0)^2\le 0$, whence $y_0=x_0=y_3$ and hence $\mathfrak{y}\in\chi$. (e) Let $\mathfrak{a}+\R \mathfrak{z}$ be a timelike line, whence $\mathfrak{z}^{\cdot 2}>0$. Hence $z_0\ne z_3$. Put $\alpha:=(a_0-a_3)/(z_3-z_0)$. Verify $\mathfrak{a}
+\alpha \mathfrak{z}\in\chi$. (f), (g) are easy to check.\qed 

An extreme example is

\begin{Exa}\label{CFLC} 
{\it The closed future light cone $\{\mathfrak{x}\in\R^4: x_0=|x|\}$ is a maximal achronal not spacelike set.} Indeed, this holds by (\ref{PAS})(g) and since $||x|-|y||\le|x-y|$ for all $x,y\in\R^3$.
\end{Exa}

\subsubsection{Causal bases}

 Commonly one regards two events $\mathfrak{x}$,  $\mathfrak{y}$ to be \textbf{causally independent} if they cannot be joined by a signal moving not faster than light. Hence two events are causally independent if and only if they are spacelike separated. In particular   lightlike related events are considered to be causally dependent. 
A familiar idea of a causal evolution  is that the \textbf{Cauchy data} given on a causally independent set determine by means of a hyperbolic PDE all  the events which can be joined by signals moving not faster than light.  The following definition of a causal base is inspired by this idea.

\begin{Def}\label{DCB}
 A spacelike set $\Sigma$ is called a  \textbf{causal base} if $\Sigma$ intersects all causal  (i.e., timelike or lightlike) lines. Roughly speaking,  the causal bases are exactly those sets of independent events,   which  determine all events of Minkowski spacetime.
\end{Def} 

\begin{Propb}\label{PCB}\\

(a) {\it Causal bases $\Sigma$ are maximal achronal and hence, in particular, maximal spacelike}.  (See (\ref{PAS})(e).)\\

(b) {\it  A causal base $\Sigma$ meets  every causal line  just once.} 
\\
\hspace*{6mm}

{\it Proof.} Let $\mathfrak{a}+\R\mathfrak{z}$ be causal and assume $\mathfrak{a}+s \mathfrak{z}$ and $\mathfrak{a}+s'\mathfrak{z}$ to be in $\Sigma$ for $s,s'\in\R$. However, these events are not spacelike separated  if they do not coincide.\qed
\\

(c) If $\Sigma$ is a causal base then  $\Sigma=\{\big(\tau(x),x\big):x\in\R^3\}$ for  a  function $\tau:\R^3\to \R$ satisfying $|\tau(x)-\tau(y)|<  |x-y|$ for $x\ne y$. (See (\ref{PAS})(i).)\\

(d)  {\it The set $\Sigma$ is a causal base if $\Sigma=\{\big(\tau(x),x\big):x\in\R^3\}$    for a  function  $\tau:\R^3\to \R$ satisfying $|\tau(x)-\tau(y)|<  |x-y|$, $x\ne y$  and $\limsup_{|x|\to \infty}|\tau(x)|/|x|<1$.} (See \cite[(44) Lemma]{C17}.) 

\end{Propb}
Obviously, every spacelike hyperplane is a causal base.  A spacelike maximal achronal set need not be a causal base, see $A$ in (\ref{SMA}). A maximal achronal set, which meets all causal lines need not  be a causal base, too, 
see $A$ in (\ref{NSCS}).

\begin{Exa}\label{SMA}  $A:=\{(\xi(x),x): x\in\R^3\}$ for $\xi(x):=\sqrt{|x|^2+1}$  is a  spacelike maximal achronal set,  which is not a causal base as it does not intersect all lightlike lines.

{\it Proof.} See  \cite[(45) Example]{C17}.\qed
\end{Exa}

 It follows an emblematic example, see sec.\,\ref{UEE}.
 
\begin{Exa}\label{NSCS}
 {\it Let  $\Delta:=\{\mathfrak{x}: x_0=0, x_3\le 0\}$, $\Gamma:= \{\mathfrak{x}: x_0=1, x_3>1\}$, $\Lambda:=\{\mathfrak{x}\in\R^4:x_0=x_3, 0<x_0\le 1\}$. Put
$P:=\Delta\cup\Gamma$, $A:=\Delta \cup \Lambda \cup \Gamma$.}
{\it Then $P$ is a maximal spacelike set, but $P$ is not a causal base.  $A$ is a maximal achronal set, but $A$ is not  spacelike  and hence not a causal base, too. Finally, $P$ meets neither all timelike  nor all lightlike lines, but $A$   intersects all causal lines and all inextendible timelike $C^1$-curves.}
\end{Exa}\\
{\it Proof.} See \cite[(46) Example, (112) Example]{C17}.  As to the final  part of the assertion note that e.g. $(0,0,0,\frac{1}{2})+\R(1,0,0,0)$ and  $(0,0,0,\frac{1}{2})+\R(1,1,0,0)$ do not meet $P$. Turning to the claim about $A$, note that $A$ equals the graph of the continuous $\tau:\R^3\to \R$, $\tau(x):=0$ if $x_3\le 0$, $:=x_3$ if $0<x_3\le 1$, and $:=1$ if $x_3 >1$. Now  let  $\mathfrak{a}+\R\mathfrak{z}$ with $\mathfrak{z}=(1,z)$. Note  $(\mathfrak{a}+\R\mathfrak{z})_0=a_0+s$.  Then $a_0+s<0\le \tau(a+sz)$ for $s<-a_0$ and $a_0+s>1\ge\tau (a+sz)$ for $s>1-a_0$. 
By Bolzano's theorem there exists $s^*$ in between with $\tau(a+s^*z)=a_0+s^*$, whence  $\mathfrak{a}+s^*\mathfrak{z}\in A$.
Similarly, $A$ is intersected by every inextendible timelike $C^1$-curve $\gamma$, since $\gamma_0$ is not bounded above nor below (\ref{AGSTT}).
\qed

Let us emphasize that $P$ is not a causal base despite being maximal spacelike. This means also that $P$ is not the subset of a causal base. However $P$ is a subset of $A$, which is maximal achronal.
As shown by  $A$, a maximal achronal set  need not be a causal base, even if it meets also all lightlike lines and all inextendible timelike $C^1$-curves.

\begin{Rem}\label{CSCB} General spacetime  theories study \textbf{Cauchy surfaces}. These are the sets which meet every inextendible timelike smooth curve 
exactly once \cite[Chapter 14, Definition 28]{O83}.  By \cite[Chapter 14, Lemma 29]{O83} they meet even all  inextendible causal smooth curves. \\
\hspace*{6mm}
 Cauchy surfaces in Minkowski space are maximal achronal by (\ref{PAS})(a),(e). In particular they meet all causal lines. Due to \cite[(2)]{CJ79}, \cite[Corollary 1]{CJ77} (see (\ref{PAS})(g),(e)) a result by Valter Moretti  (\ref{RCSVM})  shows that every maximal achronal set, which meets all lightlike lines, is a Cauchy surface. Hence one has

\begin{itemize}
\item[(a)] {\it A Cauchy surface is a maximal achronal set met by every lightlike line and vice versa.}
\item[(b)] {\it A spacelike Cauchy surface is a causal base and vice versa.}
\end{itemize}
There are Cauchy surfaces  which are not  spacelike and hence not  a causal base, see $A$ in  (\ref{NSCS}). Recall also (\ref{SMA}). The achronal not spacelike hyperplane (\ref{SNSAH}) is smooth maximal achronal but is not a Cauchy surface. This is important in view of the considerations in sec.\,\ref{MSLR} and sec.\,\ref{UEE} concerning the completeness of the class of achronal regions.

\end{Rem}

\subsubsection{Causal logic}\label{ICL}

The spacetime represented by $\R^4$ is provided with  the set  inclusion $\subset$ as partial ordering and the orthogonality relation  $\perp$ meaning \textbf{achronal separateness}, i.e. 
\begin{equation}\label{AOR}
\mathfrak{x}\perp\mathfrak{y}\quad  \Leftrightarrow\quad  \mathfrak{x}\ne\mathfrak{y} \textnormal{ and } (\mathfrak{x}-\mathfrak{y})^{\cdot 2}\le 0 
\end{equation}
These relations generate the lattice $\mathcal{C}_0$ of the $\perp$-complete subsets of $\R^4$. Here  
\begin{equation}
M^\perp:=\{\mathfrak{x\in\R^4}:\mathfrak{x}\perp\mathfrak{y}, \mathfrak{y}\in M\}
\end{equation}
 is the   $\perp$-complement of $M\subset\R^4$.  $M^{\land}:=(M^\perp)^\perp$ is called the $\perp$-completion of $M$, and
$M$ is said to be \textbf{$\perp$-complete} if $M=M^{\land}$. Orthogonality $\perp$ is Poincar\'e invariant, i.e.,  $g\cdot \mathfrak{x}\perp g\cdot \mathfrak{y}$ $\Leftrightarrow$ $ \mathfrak{x}\perp  \mathfrak{y}$ , $g\in\tilde{\mathcal{P}}$, whence $\mathcal{C}_0$  is Poincar\'e invariant.
\\
\hspace*{6mm}
 $\mathcal{C}_0$ is a complete orthomodular lattice \cite[sec.\,4]{CJ77}. All its elements are Lebesgue measurable, and $M^\perp$ is Borel if $M\subset\R^4$ is Borel \cite[sec.\,24.4]{C17}.  
 Therefore
  \begin{equation}\label{CL}
\mathcal{C}:=\{M\subset\R^4: M\; \perp\text{-complete Borel set}\}
\end{equation}
is a Poincar\'e invariant $\sigma$-complete orthomodular sublattice of $\mathcal{C}_0$.   $\mathcal{C}$ is called the \textbf{causal logic}.\\
\hspace*{6mm}
As shown in  (\ref{NRLCO}) there is no representation of $\mathcal{C}_0$. The lattice turns out to be too large, so to speak.
The restriction to $\mathcal{C}$ removes this problem and solves a technical question which arises at the construction of a representation of the causal logic, see  (\ref{BMMTM}).

Some results regarding the structure of the lattice  are reported as used  in the sequel. 
The set of \textbf{$\perp$-determinacy} of $M\subset\R^4$  is defined as
\begin{equation}\label{RDAR} 
M^{\sim}:=\{\mathfrak{x}:\forall\;\mathfrak{z} \textnormal{ with }\mathfrak{z}^{\cdot 2}>0\;\exists\, s\in\R \textnormal{ with } \mathfrak{x}+s\mathfrak{z}\in M\}
\end{equation}
It consists of  all points $\mathfrak{x}$ such that every timelike  line through $\mathfrak{x}$  meets $M$.\\
\hspace*{6mm}
 According to \cite[Corollary 1]{CJ77}, $(\R^4,\perp)$ is a complete $D$-space, which means that every $\perp$-complete set $M$ is the $\perp$-completion of any maximal achronal set contained in $M$ (recall (\ref{PAS})(c)). 
 The decisive property is
\begin{equation}\label{AADAC}
  A \text{\,  achronal set } \Rightarrow  A^\sim=A^\land
 \end{equation}
 see   (\ref{ASDC}).  Finally, if $M$ is $\perp$-complete and $A$ maximal achronal in $M$, then $M$ is Borel if and only if $A$ is Borel  \cite[Lemma 4.1]{CJ77}.\\
\hspace*{6mm}
For more details see \cite{CJ77} and   \cite[sec.\,24]{C17}.

\section{Why localization for spacetime regions}\label{WLSR}

Causality for relativistic quantum mechanics is as fundamental as apparently difficult to implement. By the most direct interpretation  of Einstein causality  the probability of localization in a \textbf{region of influence}  determined by the limiting velocity of light  $c$\footnote{$c=1$} is not less than that in the region of actual localization.   Given a spacetime region $\Delta\subset \R^4$, the spacetime region $\Delta'\subset\R^4$ is a region of influence of $\Delta$ if all \textbf{causal  lines}, 
which intersect  $\Delta$, meet $\Delta'$.
In principle the values of the probability of localization in the spacetime region $\Delta$ are assumed to be the expectation values of a localization operator, i.e., a positive operator $T(\Delta)$ assigned to $\Delta$. Hence the condition imposed by causality is 
\begin{equation}
T(\Delta)\le T(\Delta')\tag{CC}
\end{equation}
 In the following we will investigate the implications of this causality condition.
\subsection{Flat spacelike regions}\label{FR}
 Initially the regions under considerations are  the measurable subsets $\Delta$ of Euclidean space $\varepsilon:=\{0\}\times\R^3$.  
Then  $T$ is  assumed to be a measure with \textbf{normalization} $T(\varepsilon)=I$. This is heuristically quite clear for its meaning. Since $\varepsilon$ is a spacelike hyperplane, 
 by relativistic symmetry these considerations are valid for any spacelike hyperplane. 
 So in a second moment one deals with all flat spacelike regions, i.e., all measurable subsets of spacelike hyperplanes. \\
  \hspace*{6mm}
 $T$ extends to a normalized measure on every spacelike hyperplane. Indeed, let $\Xi$ be a flat spacelike region, which is contained in  each of the spacelike hyperplanes $\sigma$ and $\tau$. Let $T^\sigma$ and $T^\tau$ be the measures attributed to $\sigma$ and $\tau$, respectively. Then the expectation values of $T^\sigma(\Xi)$ as well of $T^\tau(\Xi)$ are the probabilities of localization in $\Xi$. Therefore $T^\sigma(\Xi)$ and $T^\tau(\Xi)$  coincide, and determine $T(\Xi)$. \\
 \hspace*{6mm}
Let  $\Delta$ be a region of some spacelike hyperplane $\sigma$. Let $\tau$ be any other spacelike hyperplane. We distinguish the minimal region of influence  $\Delta_\tau$ of $\Delta$ in $\tau$.
It is the set of all points  in $\tau$, which can be reached  from some point  in $\Delta$ by a signal not moving faster than light. 
Explicitly, $\Delta_\tau:=\{\mathfrak{x}\in \tau: \exists \,\mathfrak{z}\in\Delta \text{ with } (\mathfrak{x}-\mathfrak{z})^{\cdot  2}\ge 0\}$. 
So CC implies
\begin{equation}\label{CCFR} 
T(\Delta)\le T(\Delta_\tau)
\end{equation}

\begin{Lem}\label{ECC}  
The condition \emph{(\ref{CCFR})} for all $\Delta$ and $\tau$ is equivalent to the condition
\begin{equation}\label{CCFRE} 
T(\Gamma) +  T(\Lambda)\le I
\end{equation}
for all spacelike separated $\Gamma$ and $\Lambda$.
\end{Lem}\\
{\it Proof.} See \cite[(14) Lemma]{C17}.
Suppose  (\ref{CCFR}). Let $\Gamma$ be contained in the spacelike hyperplane $\sigma$. By  (\ref{CCFR}) $T(\Lambda)\le T(\Lambda_\sigma)$. Since $\Gamma$ and $\Lambda$ are spacelike separated, $\Gamma$ and $\Lambda_\sigma$ are disjoint. Hence $T(\Gamma)+T(\Lambda_\sigma)=T(\Gamma\cup\Lambda_\sigma)\le I$. This implies (\ref{CCFRE}). Conversely suppose (\ref{CCFRE}). Obviously $\Delta$ and $\tau\setminus\Delta_\tau$ are spacelike separated. Therefore, by  (\ref{CCFRE}), $I\ge T(\Delta)+T(\tau\setminus\Delta_\tau)=T(\Delta)+I-T(\Delta_\tau)$, whence (\ref{CCFR}).\qed

If $T$ is projection operator valued then  causality (\ref{CCFRE}) means  \textbf{local orthogonality}, i.e., $T(\Gamma)T(\Lambda)=0$  if $\Gamma$ and $\Lambda$ are spacelike separated.

\subsection{More spacelike regions}\label{MSLR} 
There seems to be no valid reason why localization should be restricted to spacelike \textbf{flat} regions. The apparatus, which ascertain the probability of localization for two spacelike  flat regions, obviously are suited,  in the case  that the regions are spacelike separated, to determine the probability of localization in the union of the regions, which is still spacelike but is no longer flat in general. This simple observation shows that the description of localization initiated in sec.\,\ref{FR} is not complete.\\
\hspace*{6mm}
Accordingly there is research in this direction mainly regarding  causal localizations for the massive scalar boson. R.F.\,Werner works with spacelike piecewise flat hypersurfaces provided with POM\footnote{personal communication}. Recently there is the thorough study by De C. Rosa, V. Moretti \cite{DM24} on POM on smooth spacelike Cauchy surfaces (cf.\,(\ref{CSCB})(b)). Such measures are explicitly constructed from the conserved covariant density currents, which determine causal  localizations for the massive scalar boson. Actually, in Theorem 49 they deal with the vast class, named $\mathcal{C}_M$, of smooth Cauchy surfaces not necessarily spacelike  (cf. (\ref{CSCB})(a)).
\\
\hspace*{6mm}
 Nevertheless $\mathcal{C}_M$ still suffers the same kind of incompleteness of the poorer class of spacelike hyperplanes in sec.\,\ref{FR}. Indeed, there are spacelike separated measurable $\Delta\subset S$, $\Delta'\subset S'$ with $S,S'\in \mathcal{C}_M$ such that there is no $S''\in\mathcal{C}_M$ satisfying $\Delta\cup\Delta'\subset  S''$. Let $\mathcal{A}$ be an apparatus suited to ascertain the probability of localization in $\Delta$. Let $\mathcal{A}'$ be analogous. Then due to the spatial separateness of $\Delta$ and $\Delta'$ the apparatus 
 $\mathcal{A}$ and  $\mathcal{A}'$ together ascertain the probability of localization in $\Delta\cup\Delta'$, what however does not fit the frame as no  localization operator $T(\Delta\cup\Delta')$ is assigned to  $\Delta\cup\Delta'$. 
 \\
\hspace*{6mm}
As an example take $\Delta:=\{\mathfrak{x}: x_0=0,\,x_3\le 0\}$, $\Delta':= \{\mathfrak{x}: 2x_0=x_3> 0\}$. Then $\Delta\subset \{\mathfrak{x}:x_0=0\}\in\mathcal{C}_M$, 
$\Delta'\subset\{\mathfrak{x}:  2x_0=x_3\} \in\mathcal{C}_M$, and  $\Delta\cup\Delta'$ is  spacelike maximal achronal but not smooth. Similarly, consider $\Delta$ and $\Gamma$ in (\ref{NSCS}) for $\Delta$ and 
$\Delta'$. Their union $P$ is maximal spacelike but, due to (\ref{LLI}), not contained in a smooth Cauchy surface.
 \\
\hspace*{6mm} 
 For mathematical simplicity we assume the class of regions of localization to be closed under unions of countably many mutually spacelike separated elements.

\subsection{Achronal regions and regions of the causal logic} 

For simplicity let the relativistic quantum mechanical system under consideration be a free massive particle. We imagine the particle to be represented by a timelike world line. The particle is considered to be localized in a spacetime region  $\Delta$ if its world line crosses the region $\Delta$. This simple heuristic picture of localizability   reveals an important feature pointed out by \cite[sec.\,3]{C02}. Let $\Gamma:=\R^4\setminus \Delta$ be the complement of $\Delta$. The localization of the particle in $\Delta$ obviously does not exclude its localization in $\Gamma$ as its world line may meet $\Delta$ and $\Gamma$ as well. Similarly assume that the world line of the particle meets the region $\Delta$ twice. Then there is a partition $\Delta_1$, $\Delta_2$ of $\Delta$ such that the particle is localized in $\Delta_1$ and $\Delta_2$ as well.\\
\hspace*{6mm}
 Hence, first, additivity of the probability of localization may concern  only regions which are \textbf{achronal} separated. Secondly, either (i) one considers only \textbf{achronal regions}, or  (ii) the spacetime regions under consideration are the \textbf{regions of determinacy}  of achronal sets with respect to the achronal relation. Indeed, the world line of the particle crosses an achronal region $\Delta$ if and only if it crosses the region of determinacy $\Delta^\sim$.   Due to (\ref{AADAC}) the regions of determinacy of achronal sets
 constitute the \textbf{causal logic}.
 \\
 \hspace*{6mm}
 As we will see the two approaches, henceforth  named AL (Achronal Localization) and RCL (Representation of the Causal Logic), are equivalent (\ref{AC}).
  Finally in sec.\,\ref{CALMS}  concrete Poincar\'e covariant RCL  regarding massive systems for every spin are constructed.

 \subsection{Further considerations}\label{UEE} So far  \textbf{achronal localization}, which we will study in detail, is the frame which complies most completely with  the principle of causality for quantum mechanical systems.  This is due to the fact that the class of achronal sets is complete in the sense that the maximal achronal sets are  those  sets, which are met by every timelike line just once. According to the above heuristics this means that a massive particle is localized in every maximal achronal region, whence the localization operator assigned to a maximal achronal region is the unit operator. Therefore achronal localization is normalized and satisfies the causality condition CC in the form analogous to   (\ref{ECC}) quite by definition. The original condition is verified  explicitly in (\ref{CCPOB}). Two further consequences we remind in its favor are the equivalence of AL and RCL and the orthomodularity of the causal logic. 
 \\
\hspace*{6mm}   
The class  of achronal sets comprises the spacelike sets.  However the class of spacelike sets is not complete  as there are maximal spacelike sets which are not a causal base (\ref{DCB}). An emblematic example is the set $P=\Delta\cup\Gamma$ in (\ref{NSCS}). \\
\hspace*{6mm}
 By (\ref{ECC}) $T(\Delta)+ T(\Gamma)\le I$. The causal localizations of the massive scalar boson, the Dirac electron, and the massless  Weyl fermions on spacelike hyperplanes (sec.\,\ref{FR}) actually have $T(\Delta)+ T(\Gamma)< I$. By the \textbf{high boost limit} \cite[sec.\,24]{C17}, which exists due to the causal relation (\ref{CCFR}), $T$ extends uniquely to a normalized localization on the achronal \textbf{not spacelike} hyperplane $\chi=\{\mathfrak{x}\in\R^4: x_0=x_3\}$ (\ref{SNSAH}). It assigns to the achronal not spacelike set $\Lambda\subset\chi$  in (\ref{NSCS}), by which $A=P\cup \Lambda$ is maximal achronal,  just  the localization operator $I-T(\Delta)-T(\Gamma)$ thus upholding the 
 normalization. \\
\hspace*{6mm}
The probability of localization in $\Lambda$ has two further interesting meanings. First, since $\Gamma=\tau\setminus \Delta_\tau$ for the spacelike hyperplane $\tau:=\{\mathfrak{x}\in\R^4:x_0=1\}$, one has $T(\Lambda)=T(\Delta_\tau)-T(\Delta)$. Therefore it
 indicates the amount of  probability of localization in $\tau$ for causality, which exceeds that for the localization in $\Delta$. Second, it equals the probability of localization of the system  in the strip $\{\mathfrak{x}\in\R^4:x_0=0, 0\le x_3\le 2\e^{-\rho}\}$ if it is boosted along the  third axis with rapidity $\rho\ge 0$. Exploiting further the extension of $T$ to the achronal not spacelike hyperplanes one extends this result showing the  \textbf{Lorentz contraction}  for the systems mentioned above (see \cite{C17} for the Dirac and Weyl particles).
 \\
 \hspace*{6mm}
 The foregoing considerations demonstrate that spacelike localization is not sufficient as causality CC induces  the localization in achronal hyperplanes. This fact is studied in detail in \cite{C25}.
  \\
 \hspace*{6mm}
  Moreover, just the existence of a maximal spacelike set, which is not a causal base, is the reason behind the missing orthomodularity of the lattice of the causally complete sets generated by the spacelike relation. For a proof see \cite[sec.\,11.3]{C17}. The non-orthomodularity of the lattice is already stated in \cite{C02}.
  \\
  \hspace*{6mm} 
   All these arguments clearly evidence the advantages and necessity of achronal localization.

\section{Localization in achronal  regions}\label{LAR}

 Let $\mathcal{B}^{ach}$ denote the set of Borel subsets $\Delta$ of $\R^4$, which are achronal.  Let $\mathcal{H}$ be a separable Hilbert space. 
 
\begin{Def}\label{POB} Let $T(\Delta)$ for $\Delta\in\mathcal{B}^{ach}$ be a nonnegative bounded operator on $\mathcal{H}$. Suppose  $T(\emptyset)=0$ and  $\sum_nT(\Delta_n)=I$   for every sequence $(\Delta_n)$ of mutually disjoint sets in $\mathcal{B}^{ach}$ such that $\bigcup_n\Delta_n$ is maximal achronal. Then the map $T$ is  called an \textbf{achronal localization} (AL).
\end{Def}

\begin{Rem}\label{RPOB}  Let $T$ be an AL. The sequence  $(\Delta_n)$ in (\ref{POB}) may be finite, since in this case $\Delta_1,\Delta_2,\dots,\Delta_N,\emptyset, \emptyset,\dots$  satisfies the assumptions.\\
\hspace*{6mm}
{\it $T$ is  restricted $\sigma$-additive, i.e., $\sum_nT(\Delta_n)=T(\bigcup_n\Delta_n)$ if $\Delta_n$ are mutually disjoint and $\Delta:=\bigcup_n\Delta_n$ is achronal.} Indeed, let $\Lambda$  be maximal achronal with $\Delta\subset \Lambda$. By (a), $\Lambda$ is closed and hence $\Gamma:=\Lambda\setminus  \Delta \in \mathcal{B}^{ach}$. So, recalling (b), $T(\Delta)+T(\Gamma)=I$ and  $T(\Gamma)+ \sum_nT(\Delta_n)=I$, whence the claim.\\
\hspace*{6mm}
 {\it $T$ is monotone and $T(\Delta)\le I$ for all $\Delta\in \mathcal{B}^{ach}$.} This is obvious. 
\end{Rem}

\begin{Cor}\label{POMMAS} Let $\Lambda$ be maximal achronal. Let $T_\Lambda$ denote the restriction of $T$ to the Borel sets of $\Lambda$. Then  $T_\Lambda$ is a POM on $\Lambda$.
\end{Cor}

 AL   comply with the requirements of causality. Let $\Delta,\Gamma\in \mathcal{B}^{ach}$ be achronal separated or even spacelike separated, then $T(\Delta)+T(\Gamma)=T(\Delta\cup\Gamma)\le I$. Hence CC is satisfied  in the form analogous to   (\ref{ECC}).  In (\ref{CCPOB}) CC is shown as originally stated, i.e.,
 \begin{equation}\label{CC}
  T(\Delta)\le T(\Delta_\Sigma) 
\end{equation} 
where   $\Delta_\Sigma$ is the region of influence of the region $\Delta$  in any \textbf{causal base} $\Sigma$.
 Here however arises the technical problem that $ \Delta_\Sigma$ might not be Borel although $\Delta$ and $\Sigma$ are Borel so that $T(\Delta_\Sigma)$ were not defined. The problem can be  solved extending $T$ to the set $\mathcal{A}^{ach}$
 of $T$-measurable sets. A set $Y\subset \R^4$ is said to be $T$-\textbf{measurable}, if  there are $B,C\in\mathcal{B}^{ach}$ with $B\subset Y\subset C$ and $T(C\setminus B)=0$. Clearly $Y$ is achronal.

\begin{Lem} Let $Y$ be  $T$-measurable. If $B,C\in\mathcal{B}^{ach}$ with $B\subset Y\subset C$ and $T(C\setminus B)=0$ put $T(Y):=T(B)$.  This  definition extends $T$ unambiguously to $ \mathcal{A}^{ach}$.
\end{Lem}\\
{\it Proof.} Obviously $\mathcal{B}^{ach}\subset \mathcal{A}^{ach}$. Let $Y$ be  $T$-measurable. As to the definition of $T(Y)$ suppose that also  $B'\subset Y\subset C'$ for $B',C'\in\mathcal{B}^{ach}$ with  $T(C'\setminus B')=0$. Then $B'\subset  Y\subset C$, whence $T(B')\le T(C)$ by monotony  (\ref{RPOB})(d). Since, by  (\ref{RPOB})(c), $T(C)=T(C\setminus B)+T(B)=T(B)$, one infers $T(B')\le T(B)$. Similarly  one has $T(B)\le T(B')$. Hence $T(B)=T(B')=T(Y)$. \qed\\

 Let $\mathcal{S}^{ach}$ denote the set of   achronal Suslin sets of $\R^4$.

\begin{Lem} For every AL $T$ one has $\mathcal{B}^{ach}\subset   \mathcal{S}^{ach} \subset  \mathcal{A}^{ach}$.
\end{Lem}\\
{\it Proof.} Since every Borel set of the polish space $\R^4$ is a Suslin set, $\mathcal{B}^{ach}\subset   \mathcal{S}^{ach}$ follows.
 Let $S\in \mathcal{S}^{ach}$. Choose  a
maximal achronal set $\Lambda$ with $S\subset\Lambda$.
 Let  $\mathcal{B}(\Lambda)$ and $\mathcal{S}(\Lambda)$ denote the set of Borel sets and Suslin sets of the topological subspace $\Lambda$ of $\R^4$, respectively. Note that  $\mathcal{B}(\Lambda)\subset  \mathcal{B}^{ach}$ and $\mathcal{S}(\Lambda)\subset  \mathcal{S}^{ach}$. (The former holds since $\Lambda$ is a Borel set, and the latter holds since generally the Suslin sets of a topological subspace are exactly the Suslin sets of the space, which are  contained in the subspace.)\\
\hspace*{6mm}
By (\ref{POMMAS}),
$T|_{\mathcal{B}(\Lambda)}$ is a POM  on $\Lambda$.  
Since $\Lambda$ is a metric space, by the theory of Suslin sets  
there are Borel sets $B,C\subset \Lambda$ with $B\subset S\subset C$ and $T|_{\mathcal{B}(\Lambda)}(C\setminus B)=0$. 
This means  $T(C\setminus B)=0$, whence $S$ is $T$-measurable.\qed

\begin{The}\label{CCPOB} Let $T$ be an AL.  Let $\Delta$ be achronal and $\Sigma$ be a causal base.
\begin{itemize}
\item[\emph{(a)}] If $\Delta\in  \mathcal{S}^{ach}$ then $\Delta_\Sigma$ is $T$-measurable. 
\item[\emph{(b)}] If $\Delta$ and  $\Delta_\Sigma$ are $T$-measurable then the causality condition \emph{(\ref{CC})} holds.
\end{itemize}
\end{The}

{\it Proof.} (a) $k:\Delta\times \Sigma\to \R$, $k(\mathfrak{x},\mathfrak{y}):=(\mathfrak{x}-\mathfrak{y})^{\cdot 2}$ is measurable with respect to the product $\sigma$-algebra 
$\mathcal{B}(\Delta)\otimes \mathcal{A}(\Sigma)$. Here $\mathcal{B}(\Delta)$ is the $\sigma$-algebra of the Borel sets of the subspace $\Delta$, which equals $\{B\cap \Delta: B\in\mathcal{B}^{ach}\}$, and $ \mathcal{A}(\Sigma)$ is the completion of  $\mathcal{B}(\Sigma)$ with respect to $T|_{\mathcal{B}(\Sigma)}$, which equals $\{ Y\in\mathcal{A}^{ach}: Y\subset \Sigma\}$. Therefore $K:=k^{-1}([0,\infty[ )\in \mathcal{B}(\Delta)\otimes \mathcal{A}(\Sigma)$. Let pr$_\Sigma$ be the projection of $\Delta\times \Sigma$ onto $\Sigma$.
Note pr$_\Sigma(K)=\Delta_\Sigma$. Hence by the Measurable Projection Theorem $\Delta_\Sigma$ is $T$-measurable.\\
\hspace*{6mm} (b) Let $\mathfrak{y}\in\Sigma\setminus \Delta_\Sigma$. Then by definition $(\mathfrak{y}-\mathfrak{z})^{\cdot 2} < 0$ for all $\mathfrak{z}\in\Delta$. Hence the achronal $T$-measurable sets $\Sigma\setminus \Delta_\Sigma$ and $\Delta$ are disjoint and $(\Sigma\setminus \Delta_\Sigma)\cup \Delta$ is achronal. So  (\ref{RPOB}), extended to $T$-measurable sets, yields $T(\Sigma\setminus \Delta_\Sigma)+T(\Delta)=T\big((\Sigma\setminus \Delta_\Sigma)\cup \Delta\big)\le I$ and $T(\Sigma\setminus \Delta_\Sigma)=I-T(\Delta_\Sigma)$. The causality condition (\ref{CC}) follows.
\qed\\

The result (\ref{CCPOB})(a) is  rather general as $\mathcal{S}^{ach}$ is a rich class of sets. Of course in principle it would be more satisfactory if $\mathcal{S}^{ach}$ could be replaced by $\mathcal{A}^{ach}$.
Let us recall the result \cite[(16) Lemma]{C17} concerning  the case that  $\Sigma$ is flat, i.e., a spacelike hyperplane. Accordingly, $\Delta_\Sigma$  is Lebesgue measurable for any subset $\Delta$ of a spacelike hyperplane. This result is relevant for the  positive operator-valued  localizations, which are  Poincar\'e covariant,  treated in \cite{C17}, \cite{C23}.

\section{Localization in regions of the causal logic}\label{LACR}

Recall sec.\,\ref{ICL} regarding the causal logic $\mathcal{C}=\{M\subset\R^4: M\; \perp\textrm{-complete Borel set}\}$ generated by the achronal relation, to which we will refer tacitly.

\begin{Def}\label{POC}
Let $F(M)$ for  $M\in \mathcal{C}$  be a nonnegative bounded operator on $\mathcal{H}$. Suppose $F(\emptyset)=0$, 
and $\sum_nF(M_n)=I$ for every sequence $(M_n)$ of mutually orthogonal sets in $\mathcal{C}$ 
such that 
$\bigvee_nM_n=\R^4$. Then the map $F$ is  called a representation of the causal logic  (RCL).
\end{Def}

\begin{Rems}\label{RPOC} Let $F$ be a RCL. The sequence  $(M_n)$ in (\ref{POC}) may be finite, since in this case $M_1,M_2,\dots,M_N,\emptyset, \emptyset,\dots$  satisfies the assumptions.\\
\hspace*{6mm}
 {\it $F$ is $\sigma$-orthoadditive, i.e.,  $\sum_nF(M_n)=F(\bigvee_nM_n)$ if the $M_n$ are mutually orthogonal.} Indeed, put  $M:=\bigvee_nM_n$. Then the sequence 
$M^\perp$, $M_1,M_2,\dots$ satisfies the assumptions in  (\ref{POC})  so that $F(M^\perp)+\sum_n F(M_n)=I$. Use $F(M)+F(M^\perp)=I$.
\\
\hspace*{6mm}
 {\it  If $M\subset N$ then  $F(M)+F(M^\perp \wedge N)=F(N)$, whence $F$ is monotone.}  Indeed, orthomodularity yields $N= M\vee (M^\perp \wedge N)$. Since $M \perp (M^\perp \wedge N)$,   orthoadditivity yields  the claim.

\end{Rems}

\begin{Lem}\label{CYB} Let $F$ be a RCL. Set $T(\Delta):=F(\Delta^\wedge)$ for $\Delta\in\mathcal{B}^{ach}$. Then $T$ is an AL.
\end{Lem}\\
{\it Proof.} First $T(\emptyset)=F(\emptyset)=0$. Now let $\Delta_n\in \mathcal{B}^{ach}$ be mutually disjoint with $\Delta:=\bigcup_n\Delta_n$ maximal achronal. Then  $M_n:=\Delta^{\land}_n$ is Borel. 
 Let $m\ne n$. Since  $\Delta_m\cap\Delta_n=\emptyset$ and $\Delta_m\cup\Delta_n\subset \Delta$ is achronal, one has $\Delta_m\perp\Delta_n$, whence $M_m\perp M_n$. Moreover $\emptyset=\Delta^\perp \supset \big(\bigcup_n M_n\big)^\perp$, whence  $\bigvee_nM_n=\R^4$. Hence   $(M_n)$ is a valid sequence in $\mathcal{C}$ so that $\sum_nT(\Delta_n)=\sum_nF(M_n)=I$.\qed

\begin{Lem}\label{BYC} Let  $T$ be an AL. Then there is a unique  RCL $F$ with $F(\Delta^\wedge)=T(\Delta)$ for $\Delta\in\mathcal{B}^{ach}$.
\end{Lem}\\
{\it Proof.} Let  $M\in \mathcal{C}$. Take $\Delta\subset M$  maximal achronal.  $\Delta$ is Borel and satisfies  $M=\Delta^\wedge$. Put $F(M):=A(\Delta)$.\\
\hspace*{6mm}
  $F(M)$ is well-defined. Indeed, let $\Gamma\subset M$ be maximal achronal in $M$. Then $\Gamma$ is Borel as well, and the claim is $T(\Delta)=T(\Gamma)$. Note 
  $\Delta^\perp=\Gamma^\perp=M^\perp\in\mathcal{C}$. Arguing for $M^\perp$ in place of $M$ yields a maximal achronal Borel set $\Lambda$ in $M^\perp$. So  $\Lambda^\wedge=M^\perp$. It follows $\Delta \cap \Lambda=\emptyset$,  $\Delta \cup \Lambda$ maximal achronal as well $\Gamma \cap \Lambda=\emptyset$,  $\Gamma \cup \Lambda$ maximal achronal. Therefore by (\ref{RPOB}) one has $T(\Delta)+T(\Lambda)=I$ and $T(\Gamma)+T(\Lambda)=I$, whence the claim.\\
\hspace*{6mm}
Uniqueness of $F$ is obvious. Clearly $F(\emptyset)=0$. It remains to show   $\sum_nF(M_n)=I$ for every valid sequence $(M_n)$. Recall $F(M_n)=T(\Delta_n)$ for $\Delta_n\subset M_n$ maximal achronal  in $M_n$. Then $\Delta_n$ are Borel and mutually disjoint, and $\Delta:=\bigcup_n\Delta_n$ is achronal. Moreover, $\Delta^\perp=\bigcap_n\Delta_n^\perp=\bigcap_nM_n^\perp=\big(\bigvee_nM_n\big)^{\perp}=\emptyset$, whence $\Delta^\land=\R^4$. By (\ref{AADAC}),  $\Delta$ is maximal achronal. Therefore $I=\sum_nT(\Delta_n)=\sum_nF(M_n)$.\qed

By (\ref{CYB}) and (\ref{BYC})  one has

\begin{Cor}\label{AC}  The localizations AL and RCL are equivalent descriptions of position of relativistic quantum mechanical systems.
\end{Cor}

\section{Construction of projection operator valued AL and RCL}\label{PCLL}
The method applied for the construction of RCL (and equivalently of AL) is  adopted  from Doplicher, Regge, Singer 1968 \cite{DRS68}. It produces projection valued localizations on huge carrier spaces, which contain subspaces of states representing massive  relativistic quantum mechanical systems for every spin, see  sec.\,\ref{CALMS}. The traces (compressions) of the localizations on these subspaces yield positive operator valued Poincar\'e covariant  representations of the causal logic.
\\
\hspace*{6mm}
The set  $\mathcal{T}$  of timelike lines $u:=\mathfrak{x}+\R \mathfrak{v}$, $\mathfrak{x},\mathfrak{v}\in\R^4$, $\mathfrak{v}^{\cdot2}>0$  is parametrized by $$\mathcal{T}\, \equiv\, \R^3\times O_1\;(\subset \R^6)$$
where $O_1:= \{v\in\R^3:|v|<1\}$, and $u\equiv(x,v)\equiv (0,x)+\R(1,v)$. Let $\mathcal{T}$  be   endowed  with a $\sigma$-finite measure $m$ on the Borel sets. 
Let $H_d$ be a separable Hilbert space. Then there is the canonical projection valued measure $E^{can}$ on $L_m^2(\mathcal{T}, H_d)$ with $E^{can}(\tau)$ being the multiplication operator by $1_\tau$ for  $m$-measurable $\tau\subset \mathcal{T}$.\\
\hspace*{6mm}
For every subset $M\subset\R^4$ consider the subset $\mathcal{T}_M:=\{u\in\mathcal{T}: u\cap M\ne\emptyset\}$ of $\mathcal{T}$. In general $\mathcal{T}_M$ is not $m$-measurable,  
 see \cite[(123) Example]{C17}.  However $\mathcal{T}_M$ is $m$-measurable if $M$ is Borel, see  (\ref{BMMTM}).

\begin{The}\label{CANCLL}
$F^{can}(M):=E^{can}(\mathcal{T}_M)$ for $M\in\mathcal{C}$ defines a projection operator valued \emph{RCL}. Equivalently, $T^{can}(\Delta):=E^{can}(\mathcal{T}_\Delta)$ for $\Delta\in\mathcal{B}^{ach}$ defines a projection operator valued \emph{AL}.
\end{The}\\
{\it Proof.} Due to (\ref{BMMTM}), $F^{can}(M)$ and $T^{can}(\Delta)$ are defined.  Since $\mathcal{T}_A=\mathcal{T}_{A^{\land}}$ if $A$ is achronal (use (\ref{ASDC})), $F^{can}$ and $T^{can}$ correspond with each other according to (\ref{CYB}). We show that $T^{can}$ is a AL.\\
\hspace*{6mm} 
$T^{can}(\emptyset)=0$ since $\mathcal{T}_\emptyset=\emptyset$ and $E^{can}(\emptyset)=0$. Let $\Delta_n$ be mutually disjoint achronal sets  
with
$\Delta:=\bigcup_n\Delta_n$ maximal achronal. It is easy to verify that $\mathcal{T}_{M}\cap\mathcal{T}_{N}=\emptyset$ $\Leftrightarrow$ $M\perp N$  for $M,N\subset\R^4$. Obviously
 $\bigcup_\iota \mathcal{T}_{N_\iota}=\mathcal{T}_{\cup_\iota N_\iota}$ for $N_\iota\subset\R^4$, and $\mathcal{T}_\Lambda =\mathcal{T}$ for $\Lambda$  maximal achronal.
 Therefore $\sum_nT^{can}(\Delta_n)=\sum_nE^{can}(\mathcal{T}_{\Delta_n})=E^{can}(\bigcup_n\mathcal{T}_{\Delta_n})=E^{can}(\mathcal{T}_{\cup_n\Delta_n})=E^{can}(\mathcal{T})=I$.\qed
 
 \begin{Lem} \label{MAS}
 Put $n(\Delta):=m(\mathcal{T}_\Delta)$ for $\Delta\in\mathcal{B}^{ach}$. Then $ n(\Delta)=0\,\Leftrightarrow \,T^{can}(\Delta)=0$.  
  \end{Lem}
  \\
 {\it Proof.} For $\psi\in  L_m^2(\mathcal{T}, H_d)$, $\langle \psi,T^{can}(\Delta)\psi\rangle= \langle \psi,E^{can}(\mathcal{T}_\Delta)\psi \rangle=\int_{\mathcal{T}_\Delta}||\psi(\cdot)||^2\d m$, whence the claim.\qed

\begin{Rem}\label{CMCLL}
 The result (\ref{MAS}) on $n$  remains valid for the traces of $T^{can}$   on the carrier spaces of  states of the massive systems in sec.\,\ref{CALMS}. Plainly for any trace $T$ of $T^{can}$ one has  $ n(\Delta)=0\,\Rightarrow \,T(\Delta)=0$. 
\end{Rem}

 We take a closer look to the localization  $T^{can}_\Lambda$ on a single maximal achronal region $\Lambda$.  Particularly interesting is the result (\ref{LSMAR})(d).
 \\
\hspace*{6mm} 
 The restriction $n_\Lambda$ of $n$  (\ref{MAS}) to the Borel sets of $\Lambda$ is a $\sigma$-finite measure on $\Lambda$. 
 Recall $\Lambda=\{\big(\tau(x),x\big):x\in\R^3\}$    for a $1$-Lipschitz  function $\tau:\R^3\to \R$ (\ref{PAS})(g) and the homeomorphism  $j_\Lambda:\R^3\to \Lambda$, $j_\Lambda(x):=(\tau(x),x)$ (\ref{PAS})(h). $\lambda^n$ denotes the Lebesgue measure on $\R^n$ for $n\in\N$. 
 
 \begin{Lem}\label{LSMAR}
Let $\Lambda\subset\R^4$ be maximal achronal. Let $\Delta\subset\Lambda$ be Borel, whence $\mathcal{T}_\Delta$ is $m$-measurable and the projection $\varpi(\Delta)=j_\Lambda^{-1}(\Delta)$ is Borel.
Then  
\\
\hspace*{6mm}
\emph{(a)}
 $\mathcal{T}_\Delta=\{(x-\tau(x)v,v):x\in \varpi(\Delta), v\in O_1\}$. 
\\
\hspace*{6mm}
\emph{(b)}  $k:\R^3\times O_1\to \R^3\times O_1$,  $k(x,v):=(x-\tau(x) v,v)$ is a homeomorphism. $k$ is a diffeomorphism iff $\tau$ is $C^1$.
\\
\hspace*{6mm}
\emph{(c)} 
$n_\Lambda(\Delta)=k^{-1}(m)(\varpi(\Delta)\times O_1)$ 
\\
\hspace*{6mm}
\emph{(d)} If $\tau$  is $C^1$ and $m$ is the Lebesgue measure\footnote{is short for the restriction of the Lebesgue measure $\lambda^6$}, then $n_\Lambda$ and $j_\Lambda(\lambda^3)$ are equivalent measures on $\Lambda$.
\end{Lem}\\
{\it Proof.} (a) As  $A=\{(\tau(x),x):x\in \varpi(\Delta)\}$ (\ref{PAS})(b),  $\mathcal{T}_\Delta=\{u\in\mathcal{T}: u=(\tau(x),x)+\R(1,v), \,x\in \varpi(\Delta), v\in O_1\}$, whence the claim. 
\\
\hspace*{6mm}
(b) Clearly $k$ is continuous. --- Let $k(x,v)=k(x',v')$. Then $v=v'$ and $x-\tau(x)v=x'-\tau(x')v$. For $x\ne x'$ this implies $|x-x'|<|\tau(x)-\tau(x')|\le |x-x'|$, which is a contradiction. Hence $x=x'$. Therefore $k$ is injective. --- By (a) $\mathcal{T}=\mathcal{T}_\Lambda=\{(x-\tau(x)v,v):x\in \R^3, v\in O_1\}=k(\R^3\times O_1)$, whence $k$ is surjective. --- One turns to the continuity of $k^{-1}$. Let $y_n:=k(x_n,v_n)$ tend to $y_0:=k(x_0,v_0)$ for $n\to \infty$. Clearly this implies $v_n\to v_0$.    It remains to show $x_n\to x_0$.  First assume that $(x_n)$ is unbounded.    Let $\delta<1$ such that $|v_n|\le \delta$ for all $n\ge N$. $ y_n +\tau(0)v_n=(x_n-0)-(\tau(x_n)-\tau(0))v_n$ tends to $y_0+\tau(0)v_0$ and hence is bounded.  However, $| y_n +\tau(0)v_n|\ge |x_n|-|x_n-0|\delta=|x_n|(1-\delta)$ is unbounded by assumption. This is a contradiction. So $(x_n)$ is bounded. Therefore there is  a subsequence  converging to some $x'_0$. This implies $y_0=k(x_0',v_0)$ and hence $x_0'=x_0$ as $k$ is injective.  Since this argumentation applies to every subsequence of $(x_n)$, $x_n\to x_0$ follows. The second part of the assertion in (b) is obvious.
\\
\hspace*{6mm}
(c) $k^{-1}(m)(\varpi(\Delta)\times O_1)= m(k(\varpi(\Delta)\times O_1))=m(\mathcal{T}_\Delta)=n_\Lambda(\Delta)$ by (a) and the definition of $n_\Lambda$.
\\
\hspace*{6mm}
(d) $0=j_\Lambda(\lambda^3)(\Delta)=\lambda^3(\varpi(\Delta))$ $\Leftrightarrow$ $0=m(\varpi(\Delta)\times O_1)$ $\Leftrightarrow$ $0=k^{-1}(m)(\varpi(\Delta)\times O_1)$, since $k$ is a diffeomorphism by (b). The claim holds by (c).\qed

\begin{Rem}
 In view of the localization on smooth spacelike Cauchy surfaces by De Rosa, Moretti \cite[Definition 21]{DM24} one recalls that an AL  defines on every maximal achronal set (and hence in particular on every Cauchy surface (\ref{CSCB})(a))  a  complete POM. The coherence condition by \cite{DM24} holds inherently, and the causality condition (\ref{CC}), which due to Moretti's result  (\ref{CSCB})(b)  extends GCC of \cite{DM24}, is shown to hold in (\ref{CCPOB}) without further assumptions. Actually, regarding the AL being the traces $T$ of $T^{can}$ in (\ref{CANCLL}) with $m$   the Lebesgue measure
  $$j_S(\lambda^3)(\Delta)=0 \quad \Rightarrow \quad T(\Delta)=0$$
(cf.\,(\ref{PAS})(h)) holds   for all Borel sets $\Delta$  of a maximal achronal  $C^1$-surface $S$. If $T$ is a localization of a massive system, then the implication $\Leftarrow$ holds, too. This follows from  (\ref{LSMAR})(d), (\ref{CMCLL}).

\end{Rem}

\section{Construction of covariant representations of the  causal logic for massive systems for every spin}\label{CALMS}

A representation $F$ of the causal logic is  Poincar\'e covariant, i.e., a covariant RCL,  if there is a    representation\footnote{is short for continuous unitary  representation}  $W$  of  $\tilde{\mathcal{P}}$ on $\mathcal{H}$ such that
\begin{equation}\label{PCCLL}
F(g\cdot M)=W(g)F(M)W(g)^{-1}
\end{equation}  
holds for $g\in \tilde{\mathcal{P}}$ and $M\in\mathcal{C}$.  $W$ represents the relativistic quantum mechanical system, whose position in \textbf{spacetime} is described by  $F$. The equivalent AL $T$, corresponding to $F$ according to (\ref{CYB}), is Poincar\'e covariant
satisfying (\ref{PCCLL}) with $F$ replaced by $T$. Covariance is a plain requirement of relativistic symmetry.\\
\hspace*{6mm}
We are going to construct covariant RCL for  systems with definite spin  and  mass spectrum  of positive Lebesgue measure. The main result is

\begin{The}\label{MT}
Let $W^{m\,j}$, $m>0$, $j\in\N_0/2$, denote the irreducible representation of $\tilde{\mathcal{P}}$ for mass $m>0$ and spin $j\in\N_0/2$ with carrier space $\mathcal{H}_{m\,j} $. 
 For every Borel set $M\subset ]0,\infty[$ of positive Lebesgue  measure build the direct integral representation  $$ W^{Mj}:= \int_M W^{m\,j} \d m \quad\text{on}\quad  \mathcal{H}_{Mj}:=\int_M  \mathcal{H}_{m\,j} \d m $$
 Then  there is  a  RCL  $F$ such that  $(W^{Mj}, F)$ is Poincar\'e covariant.
\end{The} \\
For the proof of (\ref{MT}) in sec.\,\ref{PMT}, the following considerations extend the construction of a particular representation of $\tilde{\mathcal{P}}$ in Doplicher, Regge, Singer 1968 \cite{DRS68}.

\subsection{Poincar\'e covariance of the RCL}\label{CCLL}

 The considerations in sec.\,\ref{PCLL} are carried on. $\tilde{\mathcal{P}}$ acts transitively on $\mathcal{T}$ in the natural way. For $u_0:=\R\mathfrak{e}$, $\mathfrak{e}:=(1,0,0,0)$ one has the stabilizer subgroup $\tilde{\mathcal{P}}_{u_0}=\R\mathfrak{e}\times SU(2)$.  Their irreducible representations are equivalent to $d^{\mu,J}(t,B):=\operatorname{e}^{\operatorname{i}\mu\, t} D^{(J)}(B)$
for $\mu\in\R$, $J\in \N_0/2$, which act on $\C^{2J+1}$. Equip $\mathcal{T}=\R^3\times O_1$ with the Lebesgue measure $m$. It is quasi invariant under $\tilde{\mathcal{P}}$. For the Radon-Nikodym derivative see (\ref{GIR}).
\\
\hspace*{6mm}
Let $d$ be a representation of $\tilde{\mathcal{P}}_{u_0}$ acting on $H_d$. By Stone's theorem $d(t,B)=\e^{\i t\mu}D(B)$, where  $\mu$ is a selfadjoint operator commuting with the representation $D$  of $SU(2)$ on $H_d$. It  induces the position  representation $W^{pos}$ (\ref{GIR}) of $\tilde{\mathcal{P}}$ acting on $L^2(\mathcal{T},H_d)$. By the imprimitivity theorem  $E^{can}$ is covariant with respect to  $W^{pos}$, i.e., $E^{can}(g\cdot \tau)= W^{pos}(g)E^{can}(\tau)W^{pos}(g)^{-1}$  for  measurable $\tau\subset \mathcal{T}$, $g\in \tilde{\mathcal{P}}$. Recall  (\ref{CANCLL}).

\begin{Pro}\label{BRCL} 
 $(W^{pos},F^{can})$ is a covariant RCL. Equally  $(W^{pos},T^{can})$  is a covariant AL.
\end{Pro}\\
{\it Proof.} Note  $g\cdot \mathcal{T}_M= \mathcal{T}_{g\cdot M}$ for  $M\subset\R^4$. Covariance  (\ref{PCCLL}) of $F^{can}$ and $T^{can}$  by $W^{pos}$ follows. \qed

For the following it suffices to restrict oneself to the irreducible $d=d^{\mu,J}$ inducing the irreducible system of imprimitivities $(W^{pos},E^{can})$. 
Henceforth $\mu>0$, $J\in\N_0/2$ are fixed. The case $\mu<0$ can be treated analogously.
\\
\hspace*{6mm}
The task is to find the invariant subspaces on which $W^{pos}$ represents a physical system determined by positive energy and a nonnegative  mass spectrum. Actually massless systems  will not occur.  In the first step one changes to the momentum representation.

\subsection{Decomposition of $W^{mom}$}\label{DWMOM}

The momentum  space representation $W^{mom}:=\mathcal{F}W^{pos}\mathcal{F}^{-1}$, where $(\mathcal{F}\psi)(p,v)=(2\pi)^{-3/2}\int \operatorname{e}^{-
 \operatorname{i}px}\psi(x,v)\operatorname{d}^3x$ for continuous $\psi$ with compact support and $W^{pos}$ is given in (\ref{GIR}), acts on $L^2(\R^3\times O_1,\C^{2J+1})$.  Explicitly one finds
\begin{equation}\label{MRI}
\big(W^{mom}(\mathfrak{a},A)\varphi\big)(p,v)=\big(A^{-1}\cdot \mathfrak{v}\big)_0^{-3/2}\,\e^{\i\mathfrak{a}\cdot\mathfrak{p}}D^{(J)}\big(R(\mathfrak{v},A)\big)\psi(A^{-1}\cdot (p,v))
\end{equation}

Here $\mathfrak{v}:=(1,v)$, $\mathfrak{p}:=(E(p,v),p)$ with $E(p,v):=pv +\mu \sqrt{1-v^2}$, the Wigner rotation  (\ref{WR}), 
and  $SL(2,\C)$ acts  on $\R^3\times O_1$ by $A\cdot (p,v):=(A\cdot p,A\ast v)$, where $A\cdot p$ denotes the spatial vector part of $A\cdot\mathfrak{p}$ and $A\ast v:=(A\cdot\mathfrak{v})_0^{-1}A\cdot v$.

\subsubsection{The invariant subspace $\mathcal{H}_\varPi$ of positive energy and nonnegative mass spectrum} 

From (\ref{MRI}) one reads off the  energy operator and the mass-squared  operator. They are the multiplication operators by $E(p,v)$ and $\gamma(p,v):=E(p,v)^2-p^2$, respectively. 

\begin{Lem}
Put $\varPi:=\{E>0,\gamma\ge 0\}$. Call $\pi$ the restriction of $\lambda^6$ to $\varPi$. Then $\mathcal{H}_\varPi\equiv L_\pi^2(\varPi,\C^{2J+1})$ is invariant under $W^{mom}$.  Let $W^\varPi$ denote the subrepresentation of 
  $W^{mom}$ on $\mathcal{H}_\varPi$.
  \end{Lem}\\
{\it Proof.}  Recall that the sign of the energy and the mass-squared operator are Casimir operators.\qed

As an  aside consider the complement $\varXi$ of $\varPi$. The question is about the physical meaning of the states in $\mathcal{H}_{\varXi}$ or even of the superpositions in 
$\mathcal{H}_{\varXi}\oplus\mathcal{H}_{\varPi}$. One may account  $\mathcal{H}_{\varXi}$  a  \textbf{virtual background} influencing the real states. For instance it could impede the system  to be strictly localized in a bounded region. Cf. the considerations in \cite[sec.\,J]{CL15}, \cite[sec.\,19]{C17}.

 The following analysis of $\varPi$ 
 is essential for the  decomposition of $W^\varPi$. Recall sec.\,\ref{CCSWR} for the canonical cross section $Q$.

\begin{Pro}\label{API} \emph{(a)}  $\varPi_m:=\{(p,v)\in\R^3\times O_1: E(p,v)=\sqrt{m^2+p^2}\}$, $m\in[0,\mu]$ are mutually disjoint $m$-null sets with union
$\bigcup_{0\le m\le\mu}\varPi_m=\varPi$. 
 \\
\hspace*{6mm} 
 \emph{(b)} Let $S_1:=\{\omega\in\R^3,|\omega|=1\}$.  For $m\in]0,\mu[$ set $\mathfrak{p}:=\big(\sqrt{m^2+p^2},p\big)$. Then
\begin{equation} 
k_m:\R^3\times S_1\to \varPi_m, \,k_m(p,\omega):=\big(p,Q(\mathfrak{p})\ast (1-m^2/\mu^2)^{1/2}\,\omega\big)
\end{equation}
is a diffeomorphism with $k_m^{-1}(p,v)=\big(p,(1-m^2/\mu^2)^{-1/2}Q(\mathfrak{p})^{-1}\ast v\big)$.
 
\hspace*{6mm} 
 \emph{(c)} Let $\varPi':=\bigcup_{0< m<\mu}\varPi_m$. Then 
\begin{equation}\label{PEMSS}
k:]0,\mu[\times\R^3\times S_1\to \varPi', \quad k(m,p,\omega):=k_m(p,\omega)
\end{equation}
is a diffeomorphism with $k^{-1}(p,v)=\big(m,k^{-1}_m(p,v)\big)$ for $m\equiv\sqrt{\gamma(p,v)}$.
\end{Pro}\\
{\it Proof.} (a) Verify $\gamma(p,v)\le \big(|p||v|+\mu\sqrt{1-v^2}\big)^2-p^2\le \mu^2$   and  hence  $\gamma(p,v)=\mu^2$ iff $p=\mu(1-v^2)^{-1/2}v$.   Check  also $\varPi=\{E(p,v)-|p|\ge 0\}$. The proof of (a) is easily completed.
\\
\hspace*{6mm} 
(b) The multiplication operator by $\gamma$ on $\mathcal{H}_\varPi$ is  invariant under $W^\varPi$. This means $\gamma(A\cdot (p,v))=\gamma(p,v)$. Hence 
$\big(A\cdot (E(p,v),p)\big)^{\cdot 2}=(E(p,v),p)^{\cdot 2}=\big(E(A\cdot (p,v)), A\cdot p\big)^{\cdot 2}$. One obtains the covariance
\begin{equation}\label{CE}
 \big(A\cdot (E(p,v),p)\big)_0=E(A\cdot (p,v))
\end{equation}
\\
\hspace*{6mm} 
 Turn to $k_m$. It is well-defined since $\omega_m:=(1-m^2/\mu^2)^{1/2}\,\omega\in O_1$ and $E(k_m(p,\omega))=E\big(Q(\mathfrak{p})\cdot 0,Q(\mathfrak{p})\ast \omega_m\big)=\big(Q(\mathfrak{p})\cdot\big(E(0,\omega_m),0\big)\big)_0$ by (\ref{CE}) for $A=Q(\mathfrak{p})$. Now $E(0,\omega_m),0)=m$, whence the last term becomes $(\mathfrak{p})_0=\sqrt{m^2+p^2}$, as claimed. The proof of (b) is easily completed.
\\
\hspace*{6mm} 
(c) follows from (b).\qed
  
 \subsubsection{Direct  integral representation over the mass spectrum}

 The Poincar\'e group $\tilde{\mathcal{P}}$ is separable and of type $I$. Hence  $W^\varPi$ is Hilbert space isomorphic with an essentially unique direct integral representation over its irreducible components, whose momentum representations on $L^2(\R^3,\C^{2j+1})$ are
 \begin{equation}\label{IMR}
  \big(W^{m,j}(\mathfrak{a},A)\varphi\big)(p)=\sqrt{\epsilon(A^{-1}\cdot p)/\epsilon(p)}\, \e^{\i \mathfrak{a}\cdot \,\mathfrak{p}} \,D^{(j)}\big(R(\mathfrak{p},A)\big)\,\varphi(A^{-1}\cdot p)
\end{equation}
for mass $m>0$, spin $j\in\N_0/2$ with $\epsilon(p):=\sqrt{m^2+p^2}$, $\mathfrak{p}:=(\epsilon(p),p)$.
\\
\hspace*{6mm} 
 In order to decompose $W^\varPi$  one applies  the  Hilbert space isomorphism
\begin{equation*}
\iota: L_\pi^2(\varPi,\C^{2J+1})\to L_\nu^2([0,\mu]\times\R^3\times S_1,\C^{2J+1})
\end{equation*}
where 
 $\nu$ is the product measure of the Lebesgue measures on $[0,\mu]$, $\R^3$ and the normalized rotational invariant measure on $S_1$. It is given by
\begin{equation}\label{CIDI}
 \iota\varphi:=\sqrt{\d \pi/ \d k(\nu)}\circ k \,\;S\, \varphi\circ k
\end{equation}
for $k$ from (\ref{PEMSS}) (neglecting the irrelevant null sets   $\{0,\mu\}\times \R^3\times S_1$ and $\varPi_0\cup \varPi_\mu$)  and $S(m,p,\omega):=D^{(J)}\big(R(\mathfrak{w},Q(\mathfrak{p})^{-1})\big)$ with  $\mathfrak{p}=\big(\sqrt{m^2+p^2},p\big)$ and $\mathfrak{w}:=\big(1,(1-m^2/\mu^2)^{1/2}\,\omega\big)$.  Note $\iota^{-1}\phi=\sqrt{\d \nu/\d k^{-1}(\pi)}\circ k^{-1}\,(S\circ k^{-1})^*\, \phi\circ k^{-1}$.

\begin{Pro}\label{DIRMS}
For $W^{[0,\mu]}:=\iota W^\varPi \iota^{-1}$ one has $\big(W^{[0,\mu]}(\mathfrak{a},A)\phi\big)(m,p,\omega)=$
\begin{equation}\label{FMD}
\text{\footnotesize{$\left(\frac{\epsilon(A^{-1}\cdot p)}{\epsilon(p)}\right)^{1/2}$}} 
\e^{\i \mathfrak{a}\cdot \mathfrak{p}} D^{(J)}\big(R(\mathfrak{p},A)\big) \phi\big(m,A^{-1}\cdot p, R(\mathfrak{p},A)^{-1}\cdot \omega\big) 
\end{equation}
For details see  \ref{DWDIR}. Hence the direct integral representation over the mass spectrum $\int_0^\mu W^{\{m\}} \d m$ of $W^\varPi$ is achieved. Here $ W^{\{m\}}$ acts on $L^2(\R^3\times S_1,\C^{2J+1})$
 and $\big(W^{\{m\}}(\mathfrak{a},A)\phi\big)(p,\omega)$ is defined by the  formula \emph{(\ref{FMD})}.  For every measurable $M\subset[0,\mu]$ of positive Lebesgue measure there is the subrepresentation  $W^M:=\int_MW^{\{m\}} \d m$ of $W^{[0,\mu]}$. Explicitly, $W^M$ acts on $ L^2(M\times\R^3\times S_1,\C^{2J+1})$ by \emph{(\ref{FMD})}. 
\end{Pro}

\subsubsection{Decomposition of the spinor space} 

One recalls that the spin $\underline{S}$ is the angular momentum, which remains of  the total angular momentum $\underline{J}$ without the orbital angular momentum $\underline{L}$ viz. at zero momentum $\underline{P}$. In case of the representation of the rotations 
 $\big(W^{\{m\}}(B)\phi\big)(p,\omega)=D^{(J)}(B)\phi(p,B^{-1}\cdot\omega)$
the carrier space of the spin, i.e., the \textbf{spinor space}, is
\begin{equation}\label{SS}
 \mathcal{S}:=L^2(S_1,\C^{2J+1})\equiv L^2(S_1)\otimes  \C^{2J+1}
 \end{equation}
which carries the tensor representation $L\otimes D^{(J)}$ with $\big(L(B)h\big)(\omega):=h(B^{-1}\cdot \omega)$.
\\
\hspace*{6mm} 
By the Peter-Weyl theorem $L=\bigoplus_{l=0,1,2,\dots}D^{(l)}$ in the sense that there is an ONB $h_{li}$, $l\in\N_0$, $i=-l,-l+1,\dots,l-1,l$, of $L^2(S_1)$ such that $L(B)h_{li}=\sum_k D^{(l)}(B)_{ki}h_{lk}$. Recall $D^{(l)} \otimes  D^{(J) }= \oplus_{k\in I(l,J)}D^{(k)}$ with $I(l,J):=\{|l-J|,|l-J|+1\dots, l+J-1,l+J\}$.
 Hence
 \begin{equation}\label{DSP}
  L\otimes D^{(J)} = \bigoplus_{j\in\N_0/2}\nu_j D^{(j)}
  \end{equation}
with multiplicities $\nu_j:=|\{l\in\N_0:  j\in I(l,J)\}|$. One has $\nu_j=2\min\{j,J\}+1$ if $j+J$ is  integer and $=0$ otherwise. In particular  all integer spins $j$ occur for $J=0$  and all half-integer spins $j$ occur for $J=1/2$. We turn to $W^M$ in (\ref{DIRMS}).

\begin{Cor}\label{ADSS}  Let $W^{Mj}:=\int_M W^{m\,j} \d m$. Then  $W^{Mj}$ acts on $L^2(M\times \R^3,\C^{2j+1})$ with  the Lebesgue measure on $M\times\R^3$ and
\\
\hspace*{6mm}  \emph{(a)} $W^{\{m\}}=\bigoplus_{j\in\N_0/2} \nu_j W^{m\,j}$ 
\\
\hspace*{6mm}  \emph{(b)} $W^M=\bigoplus_{j\in\N_0/2} \nu_j W^{Mj}$.
\end{Cor}\\
{\it Proof.}  The first claim is obvious. (b) follows from (a) as the order of sum and integral can be switched.  (a) According to (\ref{DSP}) let $V_j$ be a subspace of $L^2(S_1,\C^{2J+1})$, on which $L\otimes D^{(J)}$  is reduced to $D^{(j)}$. Let $H_i$, $i=-j,\dots, j$, be a covariant ONB of $V_j$.
\\
\hspace*{6mm} 
Then the subspace $\mathcal{V}_j:=\{\phi\in L^2(\R^3\times \mathcal{S}): \phi(p,\omega) \in V_j\} $ is invariant and reduces  $W^{\{m\}}$ to $W^{m\,j}$. Indeed, for $\phi \in \mathcal{V}_j$ there is a unique $\varphi\in L^2(\R^3,\C^{2j+1})$ such that $\phi(p,\omega)=\sum_i\varphi_i(p)H_i(\omega)$  a.e. In view of  (\ref{DIRMS}) put $R\equiv R(\mathfrak{p},A)$ and note $D^{(J)}(R)\phi(A^{-1}\cdot p,R^{-1}\cdot \omega)=\sum_i\varphi_i(A^{-1}\cdot p)D^{(J)}(R)H_i(R^{-1}\cdot\omega)=\sum_i\varphi_i(A^{-1}\cdot p)
\sum_kD^{(j)}(R)_{ki}H_k(\omega) =\sum_k\big(\sum_iD^{(j)}(R)_{ki} \varphi_i(A^{-1}\cdot p)\big)H_k(\omega)$, whence the claim.
\\
\hspace*{6mm} 
By (\ref{DSP}) $L^2(\R^3\times \mathcal{S})=\bigoplus_j\nu_j\mathcal{V}_j$ holds. This accomplishes the proof.\qed

\subsection{Proof of  (\ref{MT}) Theorem}\label{PMT}

It suffices to assume that $M$ is bounded. The general case follows building  countable orthonormal sums $\bigoplus_n(W^{M_nj},F_n)$ for mutually disjoint bounded $M_n$ and $\mu_n\ge\sup M_n$.  Hence let $\mu>\sup M$.
\\
\hspace*{6mm}
The considerations in sec.\,\ref{DWMOM} show that there is an isometry $\beta:L^2(M\times \R^3,\C^{2j+1})\to L^2(\mathcal{T},\C^{2J+1})$ satisfying $W^{Mj}=\beta^*W^{pos}\,\beta$. Then $F:=\beta^*F^{can}\,\beta$ is an RCL covariant under $W^{Mj}$. 
\\
\hspace*{6mm}
Explicitly one has $\beta=\mathcal{F}^{-1}\,\beta_\varPi \,\iota^{-1}\, \beta_M\,\beta_j$. Here, in the obvious way,  $\beta_\varPi$ embeds $L^2(\varPi,\C^{2J+1})$ in $L^2(\R^3\times O_1,\C^{2J+1})$ and $\beta_M$ embeds  $ L^2(M\times\R^3\times S_1,\C^{2J+1})$ in  $ L^2([0,\mu]\times\R^3\times S_1,\C^{2J+1})$. According to (\ref{ADSS})(b), $\beta_j$ embeds the carrier space $L^2(M\times \R^3,\C^{2j+1})$ of $W^{Mj}$ in $L^2(M\times \R^3,\mathcal{S})\equiv  L^2(M\times\R^3\times S_1,\C^{2J+1})$. For $\iota$ see (\ref{CIDI}), and the partial Fourier transformation $\mathcal{F}$ is introduced in sec.\,\ref{DWMOM}.
 This  accomplishes the proof of  (\ref{MT}).\qed

Note that the range $R$ of $\beta$ is  invariant under $W^{pos}$. The subrepresentation  $W$ of $W^{pos}$ on $R$ and the trace of $F^{can}$ on $R$ are Hilbert space isomorphic by means of  $\beta^*|_R$ with $(W^{Mj},F)$. 

\section{Final remark} Our method does not furnish a representation of the causal logic for a definite mass $m\ge 0$ system but at least  for  systems with  arbitrarily narrow  mass spectrum and definite spin. Hopefully the integration of a causal density current over the achronal regions will succeed thus constructing a covariant representation of the causal logic for the massive scalar boson. 
\\
\hspace*{6mm}
For all other relativistic quantum particles no causal density current is know so far. Regarding  the massless scalar boson probably  it does not even exist.
\\
\hspace*{6mm}
As to particle localization commonly  the measurable subsets of surfaces for Cauchy data are equated with the regions of localization. Achronal localization shows that this  identification is unjustified. Here the  surfaces for Cauchy data are the causal bases. A causal base is spacelike and hence every measurable subset is a region of localization. However  there are achronal sets which are not  spacelike and hence not a subset of a causal base. 
In  (\ref{NSCS}) the maximal achronal set $A$ contains the maximal spacelike set  $P$, which is not a causal base. While the particle is localized in $A$ anyway, the Cauchy data on $P$ do not confirm this fact.

\vspace{2cm}
\appendix

\section{Results concerning the lattice $\mathcal{C}_0$}

One sends in ahead that  by \cite[sec.\,24.4]{C17} all elements of $\mathcal{C}_0$ are Lebesgue measurable.

\begin{The}\label{NRLCO}
There are no representations of the lattice $\mathcal{C}_0$.
\end{The}\\
{\it Proof.} Assume the existence of a representation  $(W,F)$. 
For every subset $X$ of $\R^3$ put $Q(X):=F\big((\{0\}\times X)^\land\big)$. (Recall that $\perp$-complete sets are measurable.) Clearly $Q(\R^3)=I$, $Q(\emptyset)=0$. Let $X_n\subset \R^3$ be disjoint. Then $(\{0\}\times X_n)^\land$ are spacelike separated and $\big(\bigcup_n(\{0\}\times X_n)\big)^\land =\bigvee_n(\{0\}\times X_n)^\land$. Hence $Q(\bigcup_n X_n)=F\big(\big(\{0\}\times (\bigcup_nX_n)\big)^\land\big)=F\big(\big(\bigcup_n(\{0\}\times X_n)\big)^\land\big)=F\big(\bigvee_n(\{0\}\times X_n)^\land\big)=\sum_nF\big((\{0\}\times X_n)^\land\big)=\sum_nQ(X_n)$. Furthermore, for every translation $b\in\R^3$ one has $W(b)Q(X)W(b)^{-1}=W(b)F\big((\{0\}\times X)^\land\big)W(b)^{-1}=F\big((0,b)+(\{0\}\times X)^\land\big)=F\big((\{0\}\times (b+X))^\land\big)=Q(b+X)$. \\
\hspace*{6mm} 
Hence $Q$ is a translation covariant POM  on  the power set of $\R^3$. We are going to argue that such a measure does not exist. Choose any section $K\subset \R^3$ for $\R^3/\mathbb{Q}^3$, i.e., $\{b+K:b\in \mathbb{Q}^3\}$ is a countable disjoint cover of $\R^3$. Therefore $I=Q(\R^3)=\sum_{b\in \mathbb{Q}^3}Q(b+K)$. Since $Q(b+K)=W(b)Q(K)W(b)^{-1}$ it follows $Q(K)\ne 0$. Hence $\langle\varphi,Q(K)\varphi\rangle>0$ for some state $\varphi$. Then $\langle\varphi,Q(b+K)\varphi\rangle=\langle W(b)^{-1}\varphi,Q(K)W(b^{-1})\varphi\rangle \to \langle\varphi,Q(K)\varphi\rangle>0$ for $b\to 0$. Therefore there is $\delta>0$ such that $\langle\varphi,Q(b+K)\varphi\rangle\ge\frac{1}{2} \langle\varphi,Q(K)\varphi\rangle=:c$ for all $|b|\le\delta$. It follows the contradiction $1\ge\sum_{b\in\Q^3,|b|\le\delta}\langle\varphi,Q(b+K)\varphi\rangle\ge\infty \cdot c=\infty$.\qed

\begin{Lemm} \emph{\cite[(124)]{C17}}. \label{BMMTM} If $M\subset \R^4$ is Borel then  $\mathcal{T}_M$ is $m$-measurable.
\end{Lemm}\\
 {\it Proof.} Consider the map 
$ f:\R\times \R^3\times O_1\to \R^4\times O_1,\; f(\xi,x,v):=(\xi,x+\xi v,v)$
and the projection $\pi_{\R^3\times O_1}(\xi,x,v):=(x,v)$. One easily verifies
$\mathcal{T}_M\,=\,\pi_{\R^3\times O_1}\big(f^{-1}(M\times O_1)\big)$.
Now, if $M$ is Borel, then  $f^{-1}(M\times O_1)$ is Borel, whence $\mathcal{T}_M$ is measurable by the Measurable Projection Theorem. (More precisely, as $f^{-1}(M\times O_1)$ is Borel in the product space of the Suslin spaces $\R$ and $\R^3\times O_1$, the projection $\mathcal{T}_M$
is Suslin in $\R^3\times O_1$, which together with its $m$-measurable sets is a complete measurable space. Since $\R^3\times O_1$ is  metric, the assertion follows.) \qed

 \begin{Theo}  \emph{\cite[(116)]{C17}}.\label{ASDC}
The $\perp$-completion of an achronal set coincides with its set of $\perp$-determinacy.
\end{Theo}\\
{\it Proof.} (a) Let $M\subset \R^4$. We show $M^{\sim}\subset M^{\land}$. Fix $\mathfrak{x}\in M^{\sim}$. Then there is a timelike $\mathfrak{w}$ with $\mathfrak{x}+\mathfrak{w} \in M$. Hence $\big((\mathfrak{x}+\mathfrak{w})-\mathfrak{x}\big)^{\cdot 2}=\mathfrak{w}^{\cdot 2}>0$,  whence $\mathfrak{x}\not\in M^{\perp}$. Now let 
 $\mathfrak{y}\in M^{\perp}$. Then $\mathfrak{z}:=\mathfrak{x}-\mathfrak{y}\ne 0$. Assume $\mathfrak{z}^{\cdot 2}>0$. Then there is $s\in\R$ with $\mathfrak{x}-s\mathfrak{z}\in M$. Clearly, $s\ne 1$ as $\mathfrak{y}\not\in M$. It follows the contradiction $0\ge \big((\mathfrak{x}-s\mathfrak{z})-\mathfrak{y}\big)^{\cdot 2}=(\mathfrak{z}-s\mathfrak{z})^{\cdot 2}=(1-s)^2\mathfrak{z}^{\cdot 2}>0$. Thus $\mathfrak{z}^{\cdot 2}\le 0$ contrary to the assumption. This implies $\mathfrak{x}\in M^{\perp\perp}=M^{\land}$.
 \\
\hspace*{6mm}
 (b) Now  let $A$ be an achronal set. Then  $A^{\sim}\supset A^{\land}$ holds. Indeed, let $\Lambda$ be maximal achronal with $A\subset\Lambda$ by (\ref{PAS})(c). Put $B:=\Lambda\setminus A$. Then $B\perp A$, whence $B\perp A^{\land}$.
 \\
\hspace*{6mm}
 Fix $\mathfrak{x}\in A^{\land}$ and let $\mathfrak{z}^{\cdot 2}>0$. Then, by (\ref{PAS})(e), there is $s\in\R$ with $\mathfrak{x}+s\mathfrak{z}\in \Lambda$. Assume $\mathfrak{x}+s\mathfrak{z}\in B$. Then $(\mathfrak{x}+s\mathfrak{z})\perp  \mathfrak{x}$. This implies $s\ne 0$ and the contradiction $0\ge \big((\mathfrak{x}+s\mathfrak{z})-\mathfrak{x}\big)^{\cdot 2}=s^2\mathfrak{z}^{\cdot 2}>0$. Thus $\mathfrak{x}+s\mathfrak{z}\in A$. This proves $\mathfrak{x}\in A^{\sim}$.\qed\\

\section{Canonical cross section and Wigner rotation}\label{CCSWR}

 The  \textbf{position space representation} $W^{pos}$ in sec.\,\ref{CCLL} of $\tilde{\mathcal{P}}$ acting on $L^2(\mathcal{T},H_d)$ 
\begin{equation}\label{GIR}
\big(W^{pos}(g)\psi\big)(u)=\sqrt{(\operatorname{d}m^g/ \operatorname{d}m)(u)} \,d\big(R(u,g)\big)\psi(g^{-1}\cdot u)
\end{equation}
is induced by  the representation $d$  of $\tilde{\mathcal{P}}_{u_0}$ acting on $H_d$. Explicitly, for   $u\equiv(x,v)$, $\mathfrak{v}:=(1,v)$, $\mathfrak{x}:=(0,x)$ one finds $\sqrt{\frac{\operatorname{d}m^g}{ \operatorname{d}m}(u)}=\big(A^{-1}\cdot\mathfrak{v}\big)_0^{-5/2}$ and
  $d\big(R(u,g)\big)=\e^{-\i(g^{-1}\cdot\mathfrak{x})_0|(A^{-1}\cdot\mathfrak{v})_0|^{-1}\sqrt{1-v^2}\mu} D(R(\mathfrak{v},A))$ with the Wigner rotation  $R(\mathfrak{v},A)$ with respect to the canonical cross section $Q$. One recalls  (for more details see e.g.  \cite[3.3(1)]{C17}):
  
\hspace*{6mm}
(a) 
The \textbf{canonical cross section} $Q(\mathfrak{k})$ for $ \mathfrak{k}\in\R^4$ with  $\mathfrak{k}\cdot\mathfrak{k}>0$ is the unique positive $2\times 2$-matrix satisfying $Q(\mathfrak{k})\cdot (\eta m,0,0,0)=\mathfrak{k}$, where $m=\sqrt{\mathfrak{k} \cdot \mathfrak{k}}$, $\eta=\operatorname{sgn}(k_0)$.  
 This implies (i) $Q(\mathfrak{\alpha k} )=Q(\mathfrak{k})$ for $\alpha\ne 0$ and (ii) $Q(B\cdot \mathfrak{k})=BQ(\mathfrak{k})B^{-1}$ for $B\in SU(2)$.
 
\hspace*{6mm}
(b) Let $A\in SL(2,\C)$. Then 
 \begin{equation} \label{WR}
  R( \mathfrak{k},A):=Q(\mathfrak{k})^{-1}A\,Q(A^{-1}\cdot \mathfrak{k}) 
  \end{equation}
is called the \textbf{Wigner rotation} with respect to   the  canonical cross section (a).  It satisfies  (i) $R( \mathfrak{k},A)\in SU(2)$ and (ii) $R( \mathfrak{k},B)=B$ for $B\in SU(2)$.

\section{Details on the computation of $W^{[0,\mu]}$} \label{DWDIR}

The formula  (\ref{FMD})   is established using the equations in (a) and (b) below, which are verified  tacitly  applying in (a) the facts on $Q$ and $R$ reported in sec.\,\ref{CCSWR}.

\hspace*{6mm}
(a) The claim is  $D^{(J)}\big(R(\mathfrak{p},A)\big)=
S(m,p,\omega)D^{(J)} \big(R(Q(\mathfrak{p})\cdot\mathfrak{w},A)\big) S\big(m,A^{-1}\cdot p, R(\mathfrak{p},A)^{-1}\cdot\omega\big)^{-1}$. By the definition of $S$ it suffices to show
$$R(\mathfrak{p},A)=R(\mathfrak{w},Q(\mathfrak{p})^{-1})\,R(Q(\mathfrak{p})\cdot\mathfrak{w},A) R\big(R(\mathfrak{p},A)^{-1}\cdot\mathfrak{w}, Q(A^{-1}\cdot \mathfrak{p})^{-1} \big)^{-1}$$
Put $B\equiv R(\mathfrak{p},A)$.  Using (\ref{WR})  the right side becomes $Q(\mathfrak{w})^{-1} Q(\mathfrak{p})^{-1} \,Q\big(Q(\mathfrak{p})\cdot\mathfrak{w}\big)  \,Q\big(Q(\mathfrak{p})\cdot\mathfrak{w}\big)^{-1}A \, \,Q\big(A^{-1}Q(\mathfrak{p})\cdot\mathfrak{w}\big) \,R(Q(\mathfrak{p})\cdot\mathfrak{w},A) \,R\big(R(B^{-1}\cdot\mathfrak{w}, Q(A^{-1}\cdot \mathfrak{p})^{-1} \big)^{-1}=   Q(\mathfrak{w})^{-1} Q(\mathfrak{p})^{-1} A\,\,Q\big( Q(A^{-1}\cdot \mathfrak{p})B^{-1}\cdot \mathfrak{w}\big) \;Q\big( Q(A^{-1}\cdot \mathfrak{p})B^{-1}\cdot \mathfrak{w}\big)^{-1}   Q(A^{-1}\cdot \mathfrak{p}) \;Q(B^{-1}\cdot\mathfrak{w})=  Q(\mathfrak{w})^{-1} Q(\mathfrak{p})^{-1} A\,    Q(A^{-1}\cdot \mathfrak{p}) \;Q(B^{-1}\cdot\mathfrak{w}) = Q(\mathfrak{w})^{-1} B\;Q(B^{-1}\cdot\mathfrak{w}) =R\big(\mathfrak{w},B\big) =B.  $

\hspace*{6mm}
(b) Put  $\Delta(m,p.\omega,A):=\frac{d(m,p,\omega)}{d\big(m,A^{-1}\cdot p,R(\mathfrak{p},A)^{-1}\cdot \omega\big)} 
 \big(A^{-1}\cdot\mathfrak{w}\big)_0^{-3/2}$, where 
  $d:=\sqrt{\d \pi/ \d k(\nu)}\circ k $.
The claim is  $\sqrt{\epsilon(A^{-1}\cdot p)/\epsilon(p)}=\Delta(m,p.\omega,A)$. It is easy to verify that (\ref{FMD})  as it is defines a unitary transformation, cf. (\ref{IMR}). Therefore it follows 
\begin{equation*}
\int \Big(\text{\footnotesize{$\frac{\epsilon(A^{-1}\cdot p)}{\epsilon(p)}$}}-\Delta(m,p.\omega,A)^2\Big)
 \;||\phi\big(m,A^{-1}\cdot p, R(\mathfrak{p},A)^{-1}\cdot \omega\big)||^2\d \nu(m,p,\omega) =0
\end{equation*}
for all $\phi\in L^2_\nu([0,\mu]\times \R^3\times S_1,\C^{2J+1})$, whence the claim.

\section{Result on Cauchy surfaces}

The result is by  Valter  Moretti, who  agrees to the following reproduction. 

\begin{The}\label{RCSVM} Let $A$ be a maximal achronal set which intersects every lightlike line. Then A is a Cauchy surface.
\end{The}

{\it Proof.} One has to show that every inextendible timelike smooth curve $\gamma=(c,g)$ meets $A$. Obviously it is no restriction to assume  $\gamma(0)=0$, $c(s)\to \infty$ for $s\to b>0$,    and $0<\tau(0)$ 
according to (\ref{PAS})(h), (\ref{AGSTT}). Put $V:=\{\mathfrak{x}: x_0\ge |x|\}$,  $T:=\{x_0:\mathfrak{x}\in A\cap V\}$ and     $V':=\{\mathfrak{x}: x_0= |x|\}$, $T':=\{x_0:\mathfrak{x}\in A\cap V'\}$. By (\ref{AGSTT}), $\gamma(s)\in V$ for $s\ge 0$. Clearly $T\ne\emptyset$,  $T'\ne\emptyset$.
\\
\hspace*{6mm}
(a)   (\ref{RCSVM}) holds, if $\sup T<\infty$. Indeed, let $s_1>0$ with $c(s_1)>\sup T$. Put $f(s):= \tau(g(s))-c(s)$. Then $f(0)>0$, $f(s_1)< 0$, whence by continuity there is $0<s_*< s_1$ with $f(s_*)=0$. Hence $\gamma(s_*)\in A$. 
\\
\hspace*{6mm}
(b)  $\sup T<\infty$ if  $c:=\sup T'<\infty$. Indeed, assume  $\sup T=\infty$.  Then there is $\mathfrak{z}\in T$ with $z_0 \ge 2c + \tau(0)$. This implies $z_0> |z|> 0$. As $(|z|,z)\ne 0$ is lightlike  there is   $s\ge 0$ with $\mathfrak{z}':=s(|z|,z)\in A$. Hence $s|z|\le c$. Then $\mathfrak{z},\mathfrak{z}'\in A$ with $(\mathfrak{z}-\mathfrak{z}')^{\cdot 2}=(z_o-|z|)(z_o+|z|-2s|z|)>0$, which is not possible.
\\
\hspace*{6mm}
(c) Let $e\in\R^3$, $|e|=1$ and $T_e:=\{s\ge 0: s(1,e)\in A\}$. The claim is $\sup T_e<\infty$, whence by (\ref{LLI}), $T_e$ is a compact interval.   Clearly $T_e\ne\emptyset$. Now assume that $T_e$ is unbounded.
Then there are $s_n\to \infty$ with $s_n(1,e)\in A$. 
The line $(0,e)+\R(1,e)$ is lightlike, whence $\big((0,e)+s_0(1,e)\big)\in A$ for some $s_0\in\R$. Therefore $\big((s_0,(1+s_0)e)-s_n(1,e)\big)^{\cdot 2}\le 0$ for all $n$, which is not possible.    
\\
\hspace*{6mm}
(d) Put $s_e:=\inf T_e$, $s^e:=\sup T_e$.  Let $f_e(s):=\tau(se)-s$, $s\ge 0$. By continuity  $f_e(s)>0$ for $0\le s< s_e$, $f_e(s)=0$ for $ s_e\le s\le s^e$. For $s>s^e$ one has $f_e(s)<0$. Indeed, as $(\tau(se),se))$ and  $s^e(1,e)$  lie in $A$,  $ |\tau(se)-s^e|\le s-s^e$ holds. Moreover  $ \tau(se)\ne s$. This implies $\tau(se)<s$.
\\
\hspace*{6mm}
(e) The claim is $\sup T'<\infty$. Assume the converse.  Then $\sup_e s^e=\infty$.
There are $e_n\in\R^3$, $|e_n|=1$ such that  $n<s^{e_n}\to\infty$ increasingly and, due to compactness,  $e_n$ converges to some $e_*$. 
\\
\hspace*{6mm} 
Let $n_0\ge s^{e_*}$. Then there is $s^{e_*}<s_0<s^{e_n}$ for all $n\ge n_0$.  By (d), $f_{e_*}(s_0)< 0$ and  $f_{e_n}(s_0)\ge 0$. Hence 
 the contradiction  $f_{e_n}(s_0)\to f_{e_*}(s_0)\ge 0$.
\\
\hspace*{6mm} 
(f) The result follows by (a), (b), and (e).\qed\\

The following result regards every maximal achronal set.

\begin{Lem}\label{LLI} 
Let $A$ be a maximal achronal set and let $\mathfrak{x},\mathfrak{y}\in A$ be lightlike separated. Then $\mathfrak{x}+s(\mathfrak{y}-\mathfrak{x})\in A$ for $0\le s\le1$.
\end{Lem}\\
{\it Proof.} Without restriction $\mathfrak{x}=0$ and $y_0> 0$. Note $y_0=|y|$. Assume that $\tau(s_*y) \ne s_*y_0$ for some $0<s_*<1$. As $0,\mathfrak{y}, (\tau(s_*y),s_*y)$  are in  $A$, one has $(\tau(s_*y),s_*y)^{\cdot 2}\le 0$ and   $\big((\tau(s_*y),s_*y)-\mathfrak{y}\big)^{\cdot 2}\le 0$. It follows  $\tau(s_*y) < s_*y_0$ and $|\tau(s_*y)-y_0|\le(1-s_*)y_0$, which is a contradiction.\qed\\

\textbf{Acknowledgement}.  I am very grateful and indebted  to Valter Moretti for the valuable discussions clarifying  the basic concepts. \\

\textbf{Competing interests.} The author states that there is no conflict of interests.

\textbf{Data availability.} Data sharing is not applicable to this article as no datasets were generated or analyzed during the current study.

\end{document}